\newif\ifclean
\cleantrue  
\newif\ifarxiv
\arxivtrue
\ifarxiv
\documentclass{article}
\usepackage{arxiv}
\else
\documentclass[preprint,12pt]{elsarticle}
\usepackage[letterpaper]{geometry}
\fi
\usepackage[letterpaper]{geometry}
\usepackage{graphicx}
\usepackage{cite}
\usepackage{amsmath}
\usepackage{amssymb}
\usepackage{subcaption}

\usepackage{algorithm}
\usepackage{algpseudocode}
\usepackage{xfrac}
\usepackage{amsthm}
\newtheorem*{remark}{Remark}
\usepackage{hyperref}
\usepackage{placeins}
\graphicspath{{{./}}}

\usepackage{amsthm}

\usepackage{relsize}
\usepackage{upgreek}
\ifclean
\newcommand{\COMMENT}[1]{{}}
\newcommand{\QUESTION}[1]{{}}
\newcommand{\TODO}[1]{{}}
\else
\usepackage[usenames,dvipsnames]{xcolor}
\usepackage{soul}
\newcommand{\COMMENT}[1]{\textcolor{cyan}{{[ \sc{#1} ]}}\\} 
\newcommand{\QUESTION}[1]{\textcolor{cyan}{{[{QUESTION:} \it{#1} ]}}\\} 
\newcommand{\TODO}[1]{\textcolor{red}{TODO: {\bf{#1}}}\\} 

\newcommand{\red}[1]{\textcolor{red}{{#1}}}
\newcommand{\caution}{\red{\bf Draft: \today. Do not distribute.}}
\fi

\usepackage{upgreek}

\newcommand{\aref}[1]{App.\,\ref{#1}}
\newcommand{\fref}[1]{Fig.\,\ref{#1}}
\newcommand{\tref}[1]{Table\,\ref{#1}}
\newcommand{\eref}[1]{Eq.\,(\ref{#1})}

\newcommand{\sref}[1]{Sec.\!~\ref{#1}}

\newcommand{\cref}[1]{Ref.\,\cite{#1}}
\newcommand{\crefs}[1]{Refs.\,\cite{#1}}

\newcommand{\Cbb}{\mathbb{C}}

\newcommand{\ab}{\mathbf{a}}

\newcommand{\fb}{\mathbf{f}}
\newcommand{\gb}{\mathbf{g}}

\newcommand{\xb}{\mathbf{x}}

\renewcommand{\sb}{\mathbf{s}}

\newcommand{\Ab}{\mathbf{A}}

\newcommand{\Cb}{\mathbf{C}}

\newcommand{\Eb}{\mathbf{E}}

\newcommand{\Nb}{\mathbf{N}}

\newcommand{\Ib}{\mathbf{I}}

\newcommand{\Sb}{\mathbf{S}}

\newcommand{\Xb}{\mathbf{X}}

\newcommand{\Zb}{\mathbf{Z}}

\newcommand{\Lc}{\mathcal{L}}
\newcommand{\Ic}{\mathcal{I}}

\newcommand{\Zc}{\mathcal{Z}}
\newcommand{\as}{\mathsf{a}}
\newcommand{\us}{\mathsf{u}}
\newcommand{\zs}{\mathsf{z}}

\newcommand{\Ws}{\mathsf{W}}

\newcommand{\Vs}{\mathsf{V}}
\newcommand{\bs}{\mathsf{b}}

\newcommand{\chib}{{\boldsymbol{\chi}}}

\newcommand{\omegab}{{\boldsymbol{\omega}}}
\newcommand{\Omegab}{{\boldsymbol{\Omega}}}

\newcommand{\zetab}{{\boldsymbol{\zeta}}}

\newcommand{\tr}{{\operatorname{tr}}}

\newcommand{\dev}{{\operatorname{dev}}}

\newcommand{\diag}{{\operatorname{diag}}}

\newcommand{\partialb}{\boldsymbol{\partial}}

\newcommand{\NN}{\mathsf{N}\!\mathsf{N}}

\usepackage[T1]{fontenc}

\newcommand{\vargamma}{\upgamma}

\newcommand{\density}{\rho}
\newcommand{\refdensity}{\rho_0}

\newcommand{\defgrad}{\mathbf{F}}

\newcommand{\abstemperature}{T} 
\newcommand{\temperature}{\theta}
\newcommand{\heatcapacity}{c}
\newcommand{\heatflux}{\mathbf{q}}

\newcommand{\refheatflux}{\mathbf{h}}

\newcommand{\pkstress}{\mathbf{S}}

\newcommand{\stress}{{\boldsymbol{\sigma}}}

\newcommand{\conjugateforce}{{\mathbf{K}}}
\newcommand{\strain}{{\boldsymbol{\epsilon}}}
\newcommand{\lagrangestrain}{\mathbf{E}}

\newcommand{\eqps}{\varepsilon_p}
\newcommand{\doteqps}{\dot{\varepsilon}_p}

\newcommand{\internalenergy}{\epsilon}
\newcommand{\internalvariable}{\kappa}
\newcommand{\internalvariables}{\boldsymbol{\kappa}}

\newcommand{\entropy}{\eta}
\newcommand{\freeenergy}{\psi}
\newcommand{\heatgeneration}{r}
\newcommand{\flow}{\mathbf{f}}

\newcommand{\thermalconductivity}{\kappa}
\newcommand{\thermalconjugate}{\boldsymbol{\alpha}}
\newcommand{\thermalexpansiontensor}{\boldsymbol{\beta}}
\newcommand{\thermalexpansioncoefficient}{{\beta}}

\newcommand{\consistencyrate}{\dot{\gamma}}
\newcommand{\consistencyparameter}{\gamma}
\newcommand{\dissipation}{\Gamma}
\newcommand{\invariants}{\mathcal{I}}
\newcommand{\jointinvariants}{\mathcal{J}}

\newcommand{\yield}{\Upsilon}
\newcommand{\dissipationpotential}{\phi}
\newcommand{\conductionpotential}{\Lambda}
\newcommand{\stressnorm}{\varsigma}

\newcommand{\Divg}{\boldsymbol{\nabla}_{\mathbf{X}}\cdot}
\newcommand{\grad}{\boldsymbol{\nabla}_{\mathbf{x}}}
\newcommand{\Grad}{\boldsymbol{\nabla}_{\mathbf{X}}}

\newcommand{\dt}{\Delta\! t}
\newcommand{\data}{\mathsf{D}}

\newcommand{\relu}{\operatorname{relu}}
\newcommand{\rect}{\operatorname{rect}}
\newcommand{\Sup}{\operatorname{sup}}

\ifclean
\else
\usepackage{todonotes}
\renewcommand{\TODO}[1]{\todo[color=orange]{{#1}}}
\renewcommand{\QUESTION}[1]{\todo[color=yellow]{Q:{#1}}}
\fi

\title{A physics-augmented neural network framework for modeling and detecting thermo-visco-plastic behavior}
\author{
Reese E. Jones\\
Sandia National Laboratories \\
Livermore CA, USA
\And
Asghar Jadoon\\
The University of Texas at Austin \\
Austin TX, USA
\And
D. Thomas Seidl\\
Sandia National Laboratories \\
Albuquerque NM, USA
\And
Jan N. Fuhg\\
The University of Texas at Austin \\
Austin TX, USA
}

\begin{document}
\ifclean
\date{}
\else
\date{\caution}
\fi

\maketitle{}

\begin{abstract}
Although considerable attention has been devoted to the development of models for isothermal, rate-independent plasticity, many high-consequence performance assessments involve viscoplastic processes that generate substantial heat.
In addition, materials may transit from a nearly isothermal, rate-independent regime to a viscous, temperature-dependent regime during these processes, which makes modeling more challenging.
In this work, we develop a physics-augmented neural network (PANN) framework for modeling general temperature-dependent, rate-dependent inelastic processes firmly based on physical principles, including the second law of thermodynamics and coordinate equivariance.
These embedded properties are enabled by a number of architectural innovations in the structure and training of an input convex and potential-based neural ordinary differential equation framework.
The resulting neural network models are capable of representing a wide spectrum of rate- and temperature-dependence ranging from isothermal, rate-independent elastic-plastic phenomenology to rate-dependent fully viscous inelastic behavior, as we demonstrate.
We also show that the framework is capable of modeling complex microstructural inelasticity and predicting the conversion of plastic work to heating when calibrated to stress-temperature observations.
\end{abstract}

{\bf Keywords}: Thermoviscoplasticity, Physics-augmented Neural Networks, Input Convex Neural Network, Neural Ordinary Differential Equations

\section{Introduction}
The conversion of plastic work to heat and the resulting softening can lead to runaway failure in high-strain rate applications.
Taylor-Quinney \cite{taylor1931plastic} related the temperature rise to plastic work in adiabatic conditions, and this has been the basis for a multitude of engineering calculations.
Typically, an ad hoc constant is used as a coefficient of proportionality between plastic work and heating, and yet the relationship is clearly a function of thermodynamic state \cite{rittel2017dependence}.
Furthermore, temperature and rate dependence tend to be concomitant in materials, and viscosity is typically strongly temperature dependent.
So even in lower rate processes, a temperature rise can occur that significantly changes the behavior of the material;  yet, there is a lack of constitutive models that are applicable across rate-temperature regimes.

Hence, we have developed a thermodynamically consistent, unified neural network-based modeling framework that: (a) is equally applicable across rate and temperature regimes; and (b) accurately predicts the plastic work converted to heat.
Like others \cite{lemaitre1994mechanics,chaboche2001unified}, it is based on two potentials: the (Helmholtz) free energy and a dissipation potential, both of which are influenced by an inferred internal state \cite{jones2018machine}.
All necessary properties and responses can be related to these fundamental thermodynamic potentials, and all unobservable states and fluxes are inferred from data.
In the process of creating this framework, we made a number of innovations, including:
\begin{enumerate}
\item inherent satisfaction of the  dissipation inequality due to modifications of a partially input convex neural network \cite{amos2017input},
\item joint observable-hidden state invariants to allow the internal states to affect the stress in a tensorial manner,
\item a rectified dissipation potential and custom loss that penalizes lack of duality to enable  state-aware rate-independent or dependent flow,  and
\item a means of injecting prior beliefs on the phenomenology through penalties that enforce a finite elastic region and yield-flow duality.
\end{enumerate}
In the next section \sref{sec:related} we put the present developments in the context of related work.
Then, in  \sref{sec:phenomenology}, we briefly summarize phenomenology and basic theory to put the framework development in context.
Subsequently, \sref{sec:theory} provides the theory to motivate and delineate the model architecture described in \sref{sec:model}.
\sref{sec:data} provides the details of the data and training.
\sref{sec:results} demonstrates the proposed framework on three exemplars:
(a) fitting and predicting behavior across the spectrum of viscoplastic phenomenology,
(b) homogenizing the response of crystal plasticity stochastic volume elements which have no closed form, and
(c) calibration to and prediction of an emulated tension experiment with stress-temperature evolution.
We conclude, in \sref{sec:summary}, with a summary of the performance of the learning framework and avenues for future work.

\section{Related work}\label{sec:related}

The evolution of thermoplasticity theory follows two intertwined threads: thermodynamic formalisms and phenomenological constitutive models.

\paragraph{Thermodynamic foundations}
Early 20th-century continuum thermodynamics established the foundations of energy and entropy balance in deformable solids.
The Coleman–Gurtin (1967) \cite{coleman1967thermodynamics} \emph{internal state variable} (ISV) theory introduced a broad thermodynamic framework in which internal variables evolve according to general laws constrained by the second law.
The Halphen–Nguyen (1975) \cite{halphen1975materiaux} \emph{generalized standard material} (GSM) formulation specializes these ideas by requiring that dissipation derive from convex dual potentials, leading to a unified variational structure for plasticity and viscoplasticity motivated by irreversible thermodynamics \cite{onsager1931reciprocalI,onsager1931reciprocalII,prigogine1963introduction,de2013non}.
GSM was later expanded in the \emph{hyperplasticity} formulations of Houlsby \& Puzrin (2006, 2007) \cite{houlsby2006principles,houlsby2007advanced}, which appeal to degree 1 homogeneity rather than convexity.
Chaboche and Lemaitre \cite{chaboche1987continuum,lemaitre1994mechanics} developed a similar two-potential, albeit less restrictive framework than GSM, for treating damage mechanics and non-associative plasticity.
Unlike GSM, the Chaboche formulation does not necessitate duality relationships.
These thermodynamic foundations are further elaborated in comprehensive treatments such as Lubliner (2008) \cite{lubliner2008plasticity}, Simo \& Hughes (2006) \cite{simo2006computational}, Steinmann \& Runesson (2021) \cite{steinmann2021catalogue}, which synthesize continuum mechanics, thermodynamics, and computational inelasticity into unified frameworks for modeling plastic and viscoplastic solids.
In the finite-strain setting, Simo \& Miehe (1992) \cite{simo1992associative} provided one of the first associative coupled thermo-plastic models, introducing plastic entropy production and establishing a bridge between thermodynamics and computational implementation.

\paragraph{Phenomenology}
Before the development of rigorous thermodynamic formalisms, early phenomenological efforts focused primarily on reproducing observed mechanical behavior rather than ensuring thermodynamic consistency.
Classical models such as Bingham (1922) \cite{bingham1922fluidity} for viscoplastic flow, Norton (1929) \cite{norton1929creep} for creep, and Cross (1965) \cite{cross1965rheology} for non-Newtonian rheology described rate-dependent response through empirical relations without explicit reference to the first or second laws of thermodynamics.
Temperature in these models primarily affects the material resistance to flow or creep, typically by reducing viscosity or accelerating deformation through thermally activated processes.
With the rise of high-rate testing, models such as Cowper–Symonds (1957) \cite{cowper1957strain}, Hollomon (1945) \cite{hollomon1945time}, and Voce (1948) \cite{voce1948relationship} were introduced to describe rate and strain hardening in metals through algebraic or exponential hardening laws.
Later, constitutive models such as Johnson–Cook (1983) \cite{johnson1983constitutive}, Steinberg–Cochran–Guinan–Lund (1980) \cite{steinberg1980constitutive}, and Preston–Tonks–Wallace (2003) \cite{preston2003model} incorporated explicit coupling between strain rate, temperature, and hardening to capture thermal softening and adiabatic heating observed at high strain rates.
Among these, the Johnson–Cook model has gained significant attention for its simplicity, broad applicability, and ease of calibration across a range of materials and loading regimes \cite{lin2010modified,burley2018johnson, mareau2020thermodynamically, steden2021comprehensive}.
While these models successfully represented a wide range of deformation regimes, their formulations were largely phenomenological and not derived from thermodynamically consistent potential frameworks.
A comprehensive overview of these developments can be found in Jočbalis et al. (2023) \cite{jovcbalis2023comparative}.

\paragraph{Incorporation of rate effects}
While these phenomenological models successfully captured the apparent dependence of strength on strain rate and temperature, they generally treated these effects as empirical modifiers rather than as manifestations of an underlying viscous mechanism.
This distinction is important for understanding how viscous effects enter constitutive descriptions of inelasticity.
In classical elastoplasticity, the material response is decomposed into recoverable elastic strain and irreversible plastic strain, with plastic flow governed by a yield function and a rate-independent flow rule.
The stress state evolves instantaneously with loading, and plastic deformation occurs once the yield surface is reached, independent of the rate of deformation.
In contrast, viscoplasticity introduces an intrinsic time dependence through a constitutive relation between plastic strain rate and the driving stress, allowing for smooth transitions between elastic, viscoelastic, and plastic behavior.
This rate dependence represents viscous effects that account for delayed plastic flow, creep, and stress relaxation.
Models such as those by Perzyna (1971) \cite{perzyna1971thermodynamic} and Duvaut–Lions (1972) \cite{duvaut1972inequations} incorporate this behavior explicitly through \emph{overstress} formulations.
Finally, rate-dependent elastoplasticity models such as Johnson–Cook (1983) \cite{johnson1983constitutive} preserve the classical yield-surface structure but introduce strain-rate sensitivity through empirical functions that modify the yield stress.
These models capture apparent rate effects and thermal softening without explicitly modeling viscous relaxation, thereby bridging purely rate-independent elastoplasticity and fully viscoplastic formulations.
Each of these formulations has its own justification and scope.
Simplifying assumptions about rate or temperature effects make calibration easier and align with the type of data available.
They also offer practical benefits such as computational efficiency and numerical robustness, which are essential for large-scale simulations.
Empirical models like Johnson–Cook \cite{johnson1983constitutive} or Cowper–Symonds \cite{cowper1957strain} can be identified from limited tests because their parameters correspond directly to measurable quantities such as yield stress, strain rate, and temperature. These features make them broadly applicable across materials and loading regimes while remaining stable and inexpensive to implement.
More elaborate viscoplastic formulations require time-resolved data but provide a clearer physical basis for dissipation. In practice, each model balances accuracy, interpretability, and efficiency according to the data and conditions it was developed for.

\paragraph{Variational modeling approaches}
Modern formulations of finite-strain plasticity and viscoplasticity have increasingly adopted variational principles to ensure thermodynamic consistency and structural stability.
Classical kinematic decompositions, whether multiplicative \cite{lee1969elastic} or additive \cite{miehe1995theory}, define the fundamental link between elastic and inelastic deformation measures.
Building on these, the variational approach introduced by Ortiz and Stainier \cite{ortiz1999variational,stainier2010study,stainier2013variational} reformulates inelastic evolution as a sequence of incremental energy minimization problems, where both equilibrium and dissipation follow from a single extremal principle involving the free energy and dissipation potential.
This structure naturally enforces the first and second laws of thermodynamics and provides a rigorous basis for numerical algorithms through incremental variational updates \cite{weinberg2006variational,fancello2008variational,brassart2012convergence}.
Within this framework, rate-independent and rate-dependent responses appear as limits of the same governing potential, while the ratio of dissipated to inelastic power arises directly from the constitutive potentials themselves \cite{stainier2013variational}.
Extensions by Mosler and co-workers \cite{mosler2010variationally,mosler2010implementation} demonstrated the generality of this concept, encompassing a broad class of classical plasticity models and providing a foundation for coupling with thermal effects \cite{canadija2011thermomechanical}.
In parallel, the associative thermoplastic model of Simo and Miehe \cite{simo1992associative} explicitly introduced plastic entropy production, further clarifying the thermodynamic origin of dissipation.
Together, these developments unify elastoplasticity and viscoplasticity within a consistent variational and thermodynamic framework, where dissipation and energy storage are governed by complementary potentials.
Furthermore, Hackl and Fischer \cite{hackl2008relation} have shown that, in many cases, the variational principles are related to the more classical guiding principles such as \emph{maximum dissipation}.

\paragraph{Thermodynamically consistent machine learning models}
Thermodynamically consistent machine learning models of inelasticity have evolved toward frameworks that encode architectural features of ISV and GSM theories.
In these approaches, neural networks are not simply regression tools but serve as representations of constitutive potentials whose gradients yield stresses and internal variable evolution.
Recent developments have therefore focused on augmenting machine learning models to implicitly fulfill thermodynamic consistency in inelastic processes, resulting in Constitutive Artificial Neural Networks (CANNs) \cite{holthusen2024theory,boes2024accounting} and  Physics-Augmented Neural Networks (PANNs) \cite{fuhg2023modular,rosenkranz2024viscoelasticty,kalina2025physics,jones2025attention,flaschel2025convex,jadoon2025automated}, for example.
While CANNs prioritize interpretability and sparsity by directly embedding the free energy and dissipation potential into compact, structured networks, PANNs emphasize flexibility and expressivity, employing convex or monotonic neural potentials that enforce thermodynamic constraints by design \cite{holthusen2025complement}.
In our view, CANNs might be better suited for experimental data, where interpretability and parameter traceability are critical, whereas PANNs may be more appropriate for homogenized or multiscale datasets, where richer internal responses often require more expressive, high-capacity models.
However, this distinction becomes less clear once sparsification techniques are introduced.
As shown by Fuhg et al. (2024) \cite{fuhg2023extreme}, extreme sparsification of physics-augmented neural networks can yield compact and interpretable functional forms while maintaining the thermodynamic consistency.
Similarly, the viscoelastic PANN framework by Kalina et al. (2025) demonstrates that regularization and gating layers can automatically deactivate redundant Maxwell elements, effectively discovering reduced representations that retain physical meaning \cite{kalina2025physics}.
Gating and sparsification were also used by Jones and Fuhg (2025) to produce a model equally adept at representing elastic, viscoelastic, or elastic-plastic behavior \cite{jones2025attention}.

\section{Phenomenological theory} \label{sec:phenomenology}

\begin{table}[h]
\centering
\begin{tabular}{|l|c|}
\hline
\textbf{Name} & \textbf{symbol} \\
\hline
\hline
Internal energy & $\internalenergy$ \\
(Helmholtz) free energy & $\freeenergy = \internalenergy - \abstemperature \entropy$ \\
Dissipation potential   & $\dissipationpotential$ \\
Dissipation             & $\dissipation$ \\
Entropy                 & $\entropy = -\partialb_\abstemperature \freeenergy$ \\
Yield function          & $\yield$ \\
\hline
(Absolute) temperature  & $\temperature$ \\
Time                    & $t$ \\
Reference position      & $\Xb$ \\
Current position        & $\xb$ \\
Motion & $\xb = \chib(\Xb,t)$ \\
Deformation gradient    & $\defgrad = \partialb_\Xb \chib$ \\
(Right Cauchy-Green) stretch tensor & $\Cb = \defgrad^T \defgrad$ \\
(Lagrange) strain tensor & $\lagrangestrain = \tfrac{1}{2}(\Cb-\Ib)$ \\
(Second Piola-Kirchhoff) stress tensor & $\pkstress = \partialb_\lagrangestrain \freeenergy$ \\
(Reference) heat flux vector & $\refheatflux$ \\
\hline
Internal state   &  $\internalvariables$ \\
Conjugate force  &  $\conjugateforce = \partialb_{\internalvariables} \freeenergy$ \\
\hline
\end{tabular}
\caption{Notation. Note all potentials are densities.}
\label{tab:notation}
\end{table}

To provide background on phenomenology, plasticity, and the classical theory used to model it, in this section, we focus on yield and plastic flow.
We leave the discussion of elastic and temperature effects to the following sections.
Mainly, we put the classical theory in the context of the Coleman-Gurtin internal state variable (ISV) framework \cite{coleman1967thermodynamics} to aid developments in the following sections, but also briefly contrast this with Generalized Standard Material (GSM) theory \cite{halphen1975materiaux}.
We use the notation summarized in \tref{tab:notation}.

\paragraph{Classic rate-independent plasticity}
Classic plasticity theory is founded on the existence of a \emph{yield function}  \cite{mises1913mechanik,hencky1924theorie,hill1998mathematical}
\begin{equation} \label{eq:yield}
\yield(\stress,\internalvariables)  = \stressnorm(\stress)- h(\internalvariables) \ ,
\end{equation}
that demarcates elastic ($\yield < 0$) versus plastic ($\yield = 0$) behavior in stress ($\stress$) and internal variable $(\internalvariables$) space.
The yield condition is typically written in the separable form seen in \eref{eq:yield}, with an effective (norm-like) \cite{hosford1972generalized} function $\stressnorm$ on the stress $\stress$ and an effective radius $h$ dependent on internal state $\internalvariables$.
For example, the widely-used von Mises yield criterion:
\begin{equation}
\stressnorm(\stress) = \sqrt{\frac{3}{2}} \| \dev \stress \|_2 \ ,
\end{equation}
encodes the plastic incompressibility commonly observed in metals \cite{taylor1931plastic} through dependence only on the deviatoric part of the stress.
This criterion is also known as the $J_2$ yield condition since the 2-norm of the deviatoric stress is equivalent to the second Cayley-Hamilton invariant of $\sb \equiv \dev \stress$.

The internal variables $\internalvariables$ change the yield function in \eref{eq:yield} and thus the elastic region.
This dependence necessitates a prescription for their evolution.
Since plastic materials are dissipative materials, typically an ordinary differential equation known as the \emph{flow rule}
\begin{equation} \label{eq:flow_rule}
\dot{\internalvariables} = \flow(\internalvariables, \stress, ...) \
\end{equation}
is used to describe the inherent path dependence.
The \emph{maximum dissipation} postulate \cite{lubliner2008plasticity} is often invoked to guide this flow:
\begin{equation} \label{eq:max_diss}
\dot{\internalvariables} = \operatorname{argmax}_{\dot{\internalvariables}} \dissipation \ ,
\end{equation}
which is sometimes motivated by the perspective that quasi-equilibrium processes maximize entropy production, as characterized by the dissipation $\dissipation$.
A Lagrange multiplier $\consistencyrate$ allows for both the maximum dissipation \eqref{eq:max_diss}  and yield \eqref{eq:yield} conditions to be satisfied
\begin{equation}
\dot{\internalvariables} = \operatorname{argmax}_{\dot{\internalvariables}} \operatorname{min}_{\consistencyrate} \dissipation + \consistencyrate \yield \ ,
\end{equation}
based on the Karush-Kuhn-Tucker (KKT) \cite{karush1939minima,kuhn1951proceedings} optimality conditions for inequality constraints:
\begin{equation} \label{eq:kkt}
\yield \le 0,\quad \consistencyrate \ge 0,\quad \yield \consistencyrate = 0\, .
\end{equation}
The multiplier $\consistencyrate$ is often called the \emph{plastic consistency parameter} since it is determined by making the plastic flow consistent with the yield surface, $\yield=0$.
For example, with (associative) flow normal to the yield surface, it leads to:
\begin{equation} \label{eq:plastic_flow}
\dot{\internalvariables}
=  \dot{\gamma} \partialb_\stress \stressnorm
=  \dot{\gamma} \frac{\sb}{\|\sb\|}
=  \dot{\gamma} \Nb \ ,
\end{equation}
which acts as a dissipation potential with normal $\Nb$ to the level sets of $\stressnorm=\| \sb \|$.
Note the constraints in \eref{eq:kkt} are specific to rate-independent plasticity.
In addition to maximum principles and other optimality conditions, convex analysis  \cite{moreau2006application,han1999plasticity} and variational calculus \cite{ortiz1999variational,canadija2016variational,stainier2013variational,mosler2010variationally} have been brought to bear on plasticity.

Typically, the dissipation $\dissipation$ is given by the inner product of the rates of internal variables $\internalvariables$ and the conjugate forces $\conjugateforce = \partialb_{\internalvariables} \freeenergy$:
\begin{equation} \label{eq:diss_req}
\dissipation = \conjugateforce \cdot \dot{\internalvariables} \ge 0 \ .
\end{equation}
If the rate of change of the internal variables is derived from a dissipation potential $\dissipationpotential$, i.e., $\dot{\internalvariables} = \partialb_\conjugateforce \dissipationpotential$, then this second law requirement can be satisfied by construction.
For instance,
\begin{equation}
\dissipationpotential = \bigl( \sum_i |K_i|^p \bigr)^{1/p} = \| \conjugateforce  \|_p \ge 0 \ \ \text{or} \ \
\dissipationpotential
= \left( \conjugateforce \cdot \Ab(\strain, \dot{\strain}) \conjugateforce \right)^{1/2}
= \| \conjugateforce  \|_{\Ab} \ge 0 \ ,
\end{equation}
with $p\ge 1$ or a positive-definite $\Ab$ satisfies \eref{eq:diss_req}.
This relies on the positive \emph{homogeneity}:
\begin{equation}
\dissipationpotential(\alpha \conjugateforce) = \alpha^p \dissipationpotential(\conjugateforce)  \ ,
\end{equation}
of this norm-like form of the dissipation function and Euler's theorem:
\begin{equation}
p \dissipationpotential(\conjugateforce) = \partialb_\conjugateforce \dissipationpotential \cdot \conjugateforce \ .
\end{equation}
In fact, rate-dependence can be embedded in the order of the homogeneous function \cite{mcbride2018dissipation}.

Furthermore, the phenomenology of plasticity indicates that the (total) strain is not the primary deformation variable and also not conjugate to the stress during \emph{inelastic flow} ($\dot{\internalvariables} \neq \mathbf{0}$).
Hence, a \emph{plastic strain} $\strain_p$, is typically added as a state variable with flow like \eref{eq:plastic_flow}.
\begin{remark}
Nevertheless, the thermodynamic definition of stress from the (Helmholtz) free energy $\freeenergy$ leads to the equivalence:
\begin{equation*}
\partialb_{\strain_e}   \hat{\freeenergy}(\strain_e)
= \partialb_{\strain} \check{\freeenergy}(\strain, \strain_p) \partialb_{\strain_e} \strain
= \partialb_{\strain} \check{\freeenergy}(\strain, \strain_p)
\end{equation*}
given the additive decomposition $\strain = \strain_e + \strain_p$,
due to the fact that the plastic variables are fixed in the partial derivatives in both cases.
A similar result can be obtained for multiplicative finite strain \cite{lee1969elastic}  but is complicated by the introduction of an intermediate configuration and maps between the current, intermediate, and reference configurations \cite{lee1969elastic,mandel1965generalisation,mandel1973thermodynamics}.
\end{remark}

\paragraph{Classic rate-dependent plasticity}
In rate-independent plasticity, the yield function $\yield$ is strictly non-positive (elastic $\yield < 0$ and plastic $\yield = 0$).
The KKT conditions, \eref{eq:kkt}, ensure that the material state remains on the yield surface during loading, such that the yield surface $\yield(\stress,\internalvariables) = 0$ and internal state $\internalvariables$ evolve together.

Viscoplasticity is commonly modeled with an \emph{overstress} formulation \cite{norton1929creep,perzyna1971thermodynamic,duvaut1972inequations} whereby the yield function is allowed to be positive, i.e., above its rate-independent limit and relaxes to this value in the absence of additional loading.
In this case, the flow rule becomes simply an ordinary differential equation (ODE) (vs. the differential-algebraic system consisting of the ODE \eqref{eq:flow_rule} and the algebraic constraint $\yield = 0$):
\begin{equation}
\dot{\strain}_p =  \consistencyrate(\stress,\ldots) \Nb(\stress,\ldots)
\end{equation}
where the $\consistencyrate$ becomes a constitutive function (vs. a multiplier).
In these models, the flow rate generally depends on distance to the yield surface, i.e., the overstress $\langle \yield \rangle$:
\begin{equation}
\gamma = \frac{1}{\tau} \langle \yield \rangle  \ ,
\end{equation}
where $\langle x \rangle = \max(0,x)$ is the Macaulay bracket and $\tau$ is a time-scale related to the viscosity of the material in plastic flow.
This form satisfies the basic requirements: (a) embedding a timescale $\tau$ and (b) having the driving force for plastic flow that goes to zero as $\yield \to 0$.

\paragraph{Generalized standard materials}
The preceding discussion shows how the classical theory fits into the  Coleman-Gurtin \cite{coleman1967thermodynamics} internal state variable framework.
Generalized standard materials (GSM) \cite{halphen1975materiaux}, on the other hand, introduce a similar but distinct framework of internal variables and potentials.
In GSM, (some) response functions are partitioned into equilibrium, which have thermodynamic/static connections, and dissipative components.
Both the physical stress and the internal conjugate forces are additively split and connected to the free energy and dissipation potentials.
It is assumed that a (Biot) differential equation connects the free energy and dissipation potential \cite{rosenkranz2024viscoelasticty,flaschel2025convex}  and a Legendre-Fenchel dual of the dissipation potential is typically employed \cite{mcbride2018dissipation,flaschel2025convex}.
We consciously avoid these complications in the neural network Chaboche-like framework we develop and adopt.

\section{Theoretical framework} \label{sec:theory}
The focus of this work is on the dissipation due to non-equilibrium and irreversible processes, which generate heat and temperature effects.
We, like others \cite{rajagopal1998mechanics,hackl2008relation,reese1998theory}, utilize potentials to: (a) keep the number of representations at a minimum, (b) aid in embedding physical constraints, and (c) maintain self-consistency of the derived response functions.
We will describe thermomechanical principles, in particular, frame equivariance and the second law constraints, in this section and embed these constraints in the functional representations in the next section.
We write the following developments in the reference frame for convenience.

\paragraph{Balance of energy}
The balance of energy (the first law of thermodynamics) is:
\begin{equation} \label{eq:1st_law}
\density_0 \dot{\internalenergy} = \pkstress \cdot \dot{\lagrangestrain} - \Grad \cdot \refheatflux + \density_0 \heatgeneration \ ,
\end{equation}
in terms of the mass density $\density_0$, the internal energy $\internalenergy$, (second Piola-Kirchhoff) stress $\pkstress$, (Lagrange) strain $\lagrangestrain$, heat flux $\refheatflux$ and heat generation $\heatgeneration$.
After substitution of the Legendre transform between the internal energy $\internalenergy$ and the (Helmholtz) free energy $\freeenergy$,
\begin{equation} \label{eq:legendre}
\freeenergy(\lagrangestrain, \abstemperature) = \Sup_\entropy \internalenergy(\lagrangestrain, \entropy) - \entropy \abstemperature \ ,
\end{equation}
and the thermodynamic definitions  for stress $\pkstress$,
\begin{equation} \label{eq:thermo_stress}
\pkstress = \density_0 \partialb_\lagrangestrain \freeenergy \ ,
\end{equation}
and entropy $\entropy$,
\begin{equation} \label{eq:thermo_entropy}
\entropy = - \density_0 \partial_\abstemperature \freeenergy \ ,
\end{equation}
the balance energy can be written:
\begin{equation} \label{eq:entropy_production}
\density_0 \abstemperature \dot{\entropy} = \dissipation - \Grad \cdot \refheatflux + \density_0 \heatgeneration \ ,
\end{equation}
where $\abstemperature$ is the (absolute) temperature.
In \eref{eq:entropy_production}, the (internal) dissipation $\dissipation$ contains the difference between the work done (stress power) $\pkstress \cdot \dot{\lagrangestrain}$ and the change in free energy:
\begin{equation}  \label{eq:dissipation_def}
\dissipation = \pkstress \cdot \dot{\lagrangestrain}
- \dot{\freeenergy}
- \density_0 \entropy \dot{\abstemperature}  \ ,
\end{equation}
which distinguishes these sources of entropy from those directly related to heat conduction and external generation.
Note that dissipation $\dissipation$ contains an entropy-temperature rate term (the last term in \eref{eq:dissipation_def}) that does not appear in the purely mechanical-isothermal context and subtracts the change in free energy due to temperature change $\density_0 \partial_\abstemperature \freeenergy \dot{\abstemperature} = -\density_0 \entropy \dot{\abstemperature}$.

The heat equation (in terms of $\dot{\abstemperature}$) results from an expansion of the time derivative of the entropy $\entropy$ in this form of the energy balance:
\begin{equation} \label{eq:elastic_heat_equation}
\heatcapacity \, \dot{\abstemperature}
= \density_0 \abstemperature \thermalexpansiontensor \cdot \dot{\lagrangestrain}
+ \dissipation - \Grad \cdot \refheatflux + \density_0 \heatgeneration \ ,
\end{equation}
where $\heatcapacity = -\density_0 \abstemperature \partial^2_{\abstemperature^2} \freeenergy$ is equated with the heat capacity at constant strain and
\begin{equation}
\thermalexpansiontensor
\equiv \partial_\abstemperature \partialb_\lagrangestrain \freeenergy
= \partial_\abstemperature \pkstress
= - \partial_\lagrangestrain \entropy
\end{equation}
is the thermal expansion tensor.
The first term on the right-hand side of \eref{eq:elastic_heat_equation} is due to (reversible) thermoelastic effects.
Note that \eref{eq:elastic_heat_equation} assumes the free energy $\freeenergy$ depends only on $\lagrangestrain$ and $\abstemperature$.

\paragraph{Second law of thermodynamics}
The Clausius-Duhem form of the second law is
\begin{equation} \label{eq:clausius_duhem}
\refdensity \dot{\entropy} \ge -\Divg\left( \frac{\refheatflux}{\abstemperature} \right) + \frac{1}{\abstemperature} \refdensity \heatgeneration \ ,
\end{equation}
which constrains the thermodynamics of physical processes.
The definition of the dissipation \eref{eq:dissipation_def} allows us to write \eref{eq:clausius_duhem} as:
\begin{equation} \label{eq:2nd_law}
\dissipation - \frac{1}{\abstemperature} \refheatflux \cdot \Grad \abstemperature \ge 0 \ .
\end{equation}

It is sufficient (but not necessary) to assume that the Fourier inequality:
\begin{equation} \label{eq:fourier_inequality}
\frac{1}{\abstemperature} \refheatflux \cdot \Grad \abstemperature \le 0 \ ,
\end{equation}
holds independently such that a potential $\conductionpotential = \conductionpotential(\gb)$ gives the heat flux $\refheatflux = \partialb_\gb \conductionpotential$ where $\gb = - \tfrac{1}{\abstemperature} \Grad \abstemperature$.

\paragraph{Inelasticity}
Modeling plasticity and other dissipative processes  introduces internal variables $\internalvariables = \{ \internalvariable_a \}_a$
\begin{equation} \label{eq:inelastic_freeenergy}
\freeenergy = \freeenergy(\lagrangestrain, \abstemperature, \internalvariables)
\end{equation}
that quantify the change in material state over the process.
Adopting the Coleman-Gurtin internal state variable (ISV) framework \cite{coleman1967thermodynamics}, our primary modeling assumption is that we can model a wide class of material response as a thermoelastic material indexed by $\internalvariables$.
Based on equipresence \cite{coleman1967thermodynamics,truesdell2004non}, we assume these variables appear in the thermodynamic potentials.

Now, revisiting the previous statements:
\begin{itemize}
\item  the dissipation $\dissipation$, \eref{eq:dissipation}, simplifies to:
\begin{equation} \label{eq:dissipation}
\dissipation =
\underbrace{\pkstress \cdot \dot{\lagrangestrain}
- \partialb_\lagrangestrain \freeenergy \cdot \dot{\lagrangestrain} }_{=\mathbf{0}}
- \underbrace{\partialb_{\abstemperature} \freeenergy \cdot \dot{\abstemperature}
- \density_0 \entropy \dot{\abstemperature} }_{=\mathbf{0}}
- \partial_{\internalvariables} \freeenergy \cdot \dot{\internalvariables}
=  - \partial_{\internalvariables} \freeenergy \cdot \dot{\internalvariables}  \,,
\end{equation}
\item the heat equation, \eref{eq:elastic_heat_equation}:
\begin{equation} \label{eq:heat_eqn}
\heatcapacity \, \dot{\abstemperature}
= \density_0 \abstemperature \thermalexpansiontensor \cdot \dot{\lagrangestrain}
+  \underbrace{( \abstemperature \partial_\abstemperature \partial_{\internalvariables} \freeenergy  - \partial_{\internalvariables} \freeenergy ) \cdot \dot{\internalvariables} }_{ \abstemperature \thermalconjugate \cdot \dot{\internalvariables} + \dissipation}
- \Grad \cdot \refheatflux + \density_0 \heatgeneration \ ,
\end{equation}
has an additional term involving $\thermalconjugate \equiv \partial_\abstemperature \partial_{\internalvariables} \freeenergy$ that arises from the rate of entropy production,
\item the second law, \eref{eq:clausius_duhem}, reduces to:
\begin{equation} \label{eq:second_law}
\dissipation =
- \partial_{\internalvariables} \freeenergy \cdot \dot{\internalvariables} \le 0 \ ,
\end{equation}
with the Fourier inequality assumption \eref{eq:fourier_inequality}.
\end{itemize}
If $\dot{\internalvariables} = \mathbf{0}$ (isentropic conditions / no heat transfer), then the material behaves in a thermoelastic manner.

We define a conjugate force $\conjugateforce$ to the flow of internal states
\begin{equation} \label{eq:conjugate_force}
\conjugateforce \equiv \partialb_{\internalvariables} \freeenergy
\end{equation}
through the free energy $\freeenergy(\lagrangestrain,\internalvariables)$.
To complete the theory, we need to provide a prescription for the evolution of the internal state:
\begin{equation}
\dot{\internalvariables} = \flow(\conjugateforce,\underbrace{\internalvariables, \lagrangestrain, \dot{\lagrangestrain}, \abstemperature}_{\Zb}) \ ,
\end{equation}
which is only constrained by the second law.
However, it is convenient to associate it with a partially convex potential $\dissipationpotential = \dissipationpotential(\conjugateforce; \Zb)$:
\begin{equation} \label{eq:flow}
\dot{\internalvariables} = \fb = \partialb_\conjugateforce \dissipationpotential
\end{equation}
so that \eref{eq:dissipation} is satisfied:
\begin{equation} \label{eq:internal_dissipation}
\dissipation
= -\partialb_{\internalvariables} \freeenergy \cdot \dot{\internalvariables}
= -\conjugateforce \cdot \partialb_{\conjugateforce} \dissipationpotential \le 0
\end{equation}
by construction \footnote{Many authors, e.g., \cref{flaschel2023automated}, that use the GSM framework omit the dependence on rate in the dissipation potential.}.
This dependence on a dissipation potential moves the formulation closer to that of Chaboche \cite{chaboche1987continuum,chaboche2001unified}.
Phenomenologically, the flow of internal state affects hardening and related material responses.

\paragraph{Invariance}
Since the modeling framework is based solely on scalar-valued potentials, namely the free energy $\freeenergy$ and the dissipation potential $\dissipationpotential$, it suffices to make all their arguments scalar invariants.
Ten invariants characterize the coordinate-free dependencies of two symmetric, second-order tensors:
\begin{equation}
\begin{array}{llllll}
\multicolumn{2}{c}{\overbrace{\hspace{1in}}^{\Ic_{\Cb}}} & & & & \\
I_1 &= \tr \Cb  -3 &\quad I_4 &= \tr \Cb \dot{\Cb}   &\quad I_8 &= \tr \dot{\Cb} \\
I_2 &= \tfrac{1}{2} ( \tr \Cb^*-3 ) &\quad I_5 &= \tr \Cb^2 \dot{\Cb} &\quad I_9 &= \tr \dot{\Cb}^2  \\
I_3 &= \det \Cb -1 &\quad I_6 &= \tr \Cb \dot{\Cb}^2 &\quad I_{10} &= \tr \dot{\Cb}^3 \\
&              &\quad I_7 &= \tr \Cb^2 \dot{\Cb}^2 & \\
\multicolumn{4}{c}{\underbrace{\hspace{2in}}_{\Ic_{\Cb,\dot{\Cb}}}} & & \\
\end{array}
\end{equation}
however, since $\Cb$ and $\dot{\Cb}$ are related by a time-derivative, we can reduce the set of invariants, see the appendix of \cref{jones2022neural} for a proof.
Here we take $\Ic_{\Cb} = \{ I_1, I_2, I_3 \}$ so that the free energy reduces to
$ \freeenergy = \freeenergy(\Ic_{\Cb}, \internalvariables, \abstemperature) $
and $\Ic_{\Cb,\dot{\Cb}} = \{ I_a\}_{a=1}^{7}$ so that
$ \dissipationpotential = \dissipationpotential(\conjugateforce, \Ic_{\Cb,\dot{\Cb}}, \abstemperature) $.
Note that
(a) $\Cb^* = \det(\Cb) \Cb^{-1}$ denotes the adjugate of $\Cb$, and
(b) we employ the Cayley-Hamilton primary invariants (vs. the Landau/power invariants).
Also, we shift and scale the usual invariants such that they are zero and have time derivatives equal to $\mathbf{0}$ in the reference configuration $(\Cb = \Ib, \dot{\Cb} = \mathbf{0}$):
\begin{alignat}{3}
I_1 &=& \tr \Cb-3                  &\Rightarrow\quad& \partialb_\Cb I_1 &= \Ib \\
I_1 &=& \tfrac{1}{2} (\tr \Cb^*-3) &\Rightarrow\quad& \partialb_\Cb I_2 &= \tfrac{1}{2} ( \tr(\Cb) \Ib - \Cb)  \\
I_3 &=& \det \Cb -1                &\Rightarrow\quad& \partialb_\Cb I_2 &= \Cb^*
\end{alignat}
given the identity $I_2 = \tr \Cb^* = \tfrac{1}{2}( \tr^2 \Cb - \tr \Cb^2 )$.
This allows all invariants to be treated uniformly.
Furthermore, we take $\abstemperature$ as an additional (scalar) invariant.

In previous work \cite{jones2022neural,jones2025attention}, we assumed that $\internalvariables$ consists of scalar invariants.
Here, we adopt a more expressive representation that allows for joint invariants $\jointinvariants= \{ J_a(\Cb,\dot{\Cb}, \internalvariables)\}_{a=1}^{N_{\jointinvariants}}$ between the internal state and the observable strain (and strain rate).
This also prevents a rank collapse of the tensor basis at high symmetry states.
In particular, by mapping the first 6 internal states to the components of a symmetric second-order reference tensor $\zetab$:
\begin{equation}
\internalvariables = ( \underbrace{\kappa_1, \kappa_2, \ldots, \kappa_6}_{\zetab}, \kappa_7, \ldots, \kappa_{N_{\internalvariables}})
\end{equation}
we construct $J_{1} = \tr( \Cb \zetab )$.

The final arguments of the free energy are:
\begin{equation} \label{eq:invariant_freeenergy}
\freeenergy =  \hat{\freeenergy}(\invariants_\Cb,\jointinvariants; \, \internalvariables,\abstemperature) \ .
\end{equation}
Hence, $\jointinvariants$ influence both
the stress $\pkstress$ and the conjugate forces $\conjugateforce$
\begin{eqnarray}
\pkstress         &=& 2 \partialb_{\Cb} \freeenergy
= 2 \sum_{a\in{1,3}} \partialb_{I_a}    \freeenergy \, \partialb_{\Cb}  I_a
+ 2 \sum_{ a \in {1,N_{\jointinvariants}}} \partialb_{J_a} \freeenergy \, \partialb_{\Cb}  J_a \, , \\
\conjugateforce &=& \partialb_{\internalvariables} \freeenergy
= \underline{\partialb}_{\internalvariables} \freeenergy
+  \sum_{a \in {1,N_{\jointinvariants}}} \partialb_{J_a}   \freeenergy \, \partialb_{\internalvariables}  J_a \, .
\label{eq:K}
\end{eqnarray}
Note in \eref{eq:K}, $\underline{\partialb}_{\internalvariables} \freeenergy$ denotes the partial derivative with respect to the explicit dependence of $\freeenergy$ on $\internalvariables$.

The dissipation potential holds the explicit rate dependence:
\begin{equation} \label{eq:invariant_dissipation_potential}
\dissipationpotential = \dissipationpotential(\conjugateforce; \, \underbrace{ \internalvariables, \Ic_{\Cb,\dot{\Cb}}, \abstemperature}_{\Zc}) \ ,
\end{equation}
where the arbitrary length $\conjugateforce$ is treated in this context as a collection of scalar invariants (similar to the components of a reference tensor).
As stated before, the dissipation potential $\dissipationpotential$ engenders the flow of internal states:
\begin{equation}
\dot{\internalvariables} = \partialb_{\conjugateforce} \dissipationpotential(\conjugateforce; \Zc) \ ,
\end{equation}
Both the free energy \eqref{eq:invariant_freeenergy} and the dissipation potential \eqref{eq:invariant_dissipation_potential} are convex in their primary inputs, $\Cb$ and $\conjugateforce$, respectively.

\section{Model} \label{sec:model}
We need a representation for: (a) the free energy $\freeenergy$ \eqref{eq:invariant_freeenergy}, which provides the stress \eqref{eq:thermo_stress} and internal forces \eqref{eq:conjugate_force}, and (b) the dissipation potential $\dissipationpotential$ \eqref{eq:invariant_dissipation_potential}, which provides the flow of the internal variables \eqref{eq:flow}.

In particular, the free energy has the requirements:
\begin{itemize}
\item Convex in the invariants $I_1$, $I_2$ and $J$ of the deformation gradient and monotonically increasing in $I_1$, $I_2$ \cite{schroder2003invariant}, which does not imply the entire model is polyconvex \cite{ball1976convexity},
\item Concave in the temperature $\abstemperature$ to ensure a positive heat capacity \cite{fuhg2024polyconvex},
\item A normalized gradient in the reference state ( $\Cb = \Ib$, $\internalvariables = \mathbf{0}$, and $\abstemperature = \abstemperature_0$) to ensure that $\pkstress = \mathbf{0}$ in this state.
\end{itemize}
Likewise, the dissipation potential has the requirements:
\begin{itemize}
\item Convex in conjugate force $\conjugateforce$ to embed the dissipation constraint \eqref{eq:diss_req},
\item Normalized gradient $\partialb_{\conjugateforce} \dissipationpotential$, and hence no flow, at $\conjugateforce = \mathbf{0}$ for all states $\Zb$.
\end{itemize}

\subsection{Physics-augmented neural network architecture} \label{sec:arch}

The convexity requirements are satisfied by a \emph{centered} input convex neural network (ICNN), which is a slightly modified version of the partially convex neural network \cite{amos2017input} presented in \cref{jadoon2025inverse}:
\begin{equation}
\varphi(\conjugateforce; \Zb) = \NN(\conjugateforce; \Zb)
\end{equation}
where $\NN$ is composed of this stack of operations:
\begin{equation}
\boxed{
\begin{aligned}[c]
\textbf{convex input}    &\rightarrow&& \us_0  \nonumber\\
\textbf{nonconvex input} &\rightarrow&& \zs_0  \nonumber\\
&\ldots&&  \nonumber\\
&\ldots&&  \nonumber\\
\zs_{k+1}        &\leftarrow&& \vartheta_a (      \Vs_k \zs_k    \hspace{0.52in} + \as_k)  \nonumber\\
\us_{k+1}        &\leftarrow&& \vartheta_c ( \Vs\!\!\!\Vs_k \zs_k + \Ws_k \us_k + \bs_k)   \nonumber\\
&\ldots&&  \nonumber\\
\textbf{output}  &\leftarrow&& \vartheta(\Vs\!\!\!\Vs_n \zs_n + \Ws_n \us_n + \bs_n) \nonumber
\end{aligned}
}
\end{equation}
Here, $\Vs_k$, $\Vs\!\!\!\Vs_k$, $\Ws_k$ are weight matrices; $\as_k$, $\bs_k$ are bias vectors, and $\vartheta_a$, $\vartheta_c$, and $\vartheta$ are activation functions.
The weights $\Ws_k$ are required to be positive, and $\vartheta_c$ needs to be convex and monotonically non-decreasing \cite{jadoon2025inverse}.
The output is centered by:
\begin{equation}
\check{\varphi} = \NN(\conjugateforce; \Zb) - \partialb_{\conjugateforce} \NN(\mathbf{0}; \Zb)\cdot\conjugateforce \ ,
\end{equation}
and can also be referenced to zero by:
\begin{equation}
\hat{\varphi} = \NN(\conjugateforce; \Zb) - \partialb_{\conjugateforce} \NN(\mathbf{0}; \Zb)\cdot\conjugateforce - \NN(\mathbf{0}; \mathbf{0}) \ .
\end{equation}
Here we employ a convex free energy (as opposed to a polyconvex $\freeenergy$) due to evidence that this form enables better representation of SVE data \cite{jadoon2025inverse} (note a polyconvex free energy is not precluded but our framework).%
\footnote{ Other works \cite{kalina2022automated,fuhg2024polyconvex} adopt a referencing term of the form $- \partialb_{\lagrangestrain} \NN(\mathbf{0},\internalvariables,\abstemperature_0) (J-1)$; this only constrains the free energy potential along $J=1$ for polyconvexity.}

\begin{figure}
\centering
\includegraphics[width=0.65\linewidth]{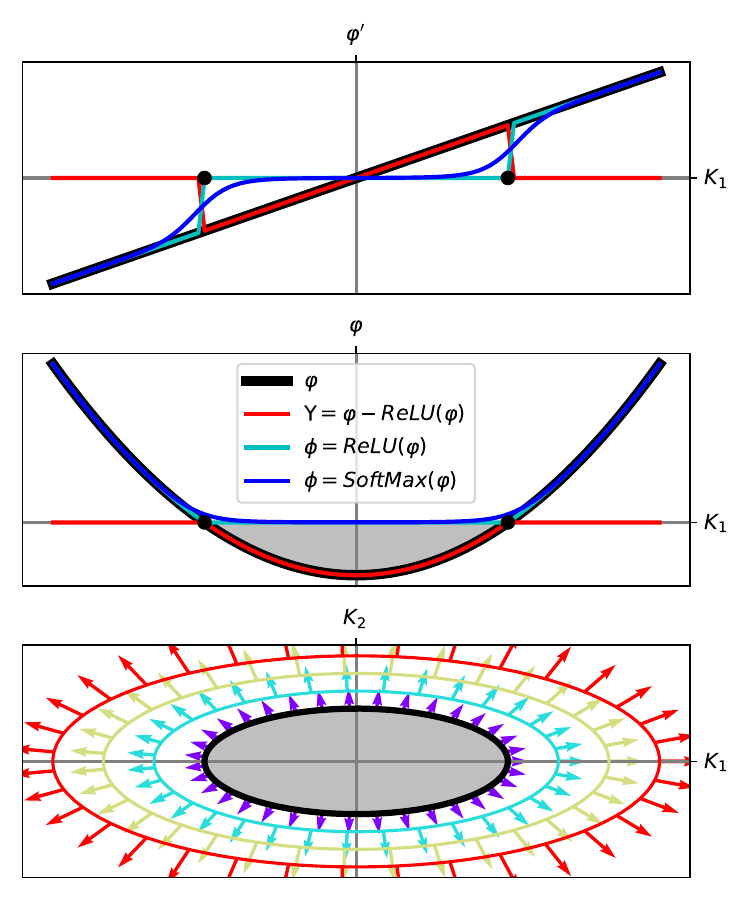}
\caption{Schematic of dissipation potential.
Gray is the elastic region with no flow of internal variables.
The effect of different rectifiers, ReLU and SoftPlus, on the dissipation potential are shown in blue and cyan, respectively.}
\label{fig:dissipation_potential}
\end{figure}

\begin{figure}
\centering
\includegraphics[width=0.55\linewidth]{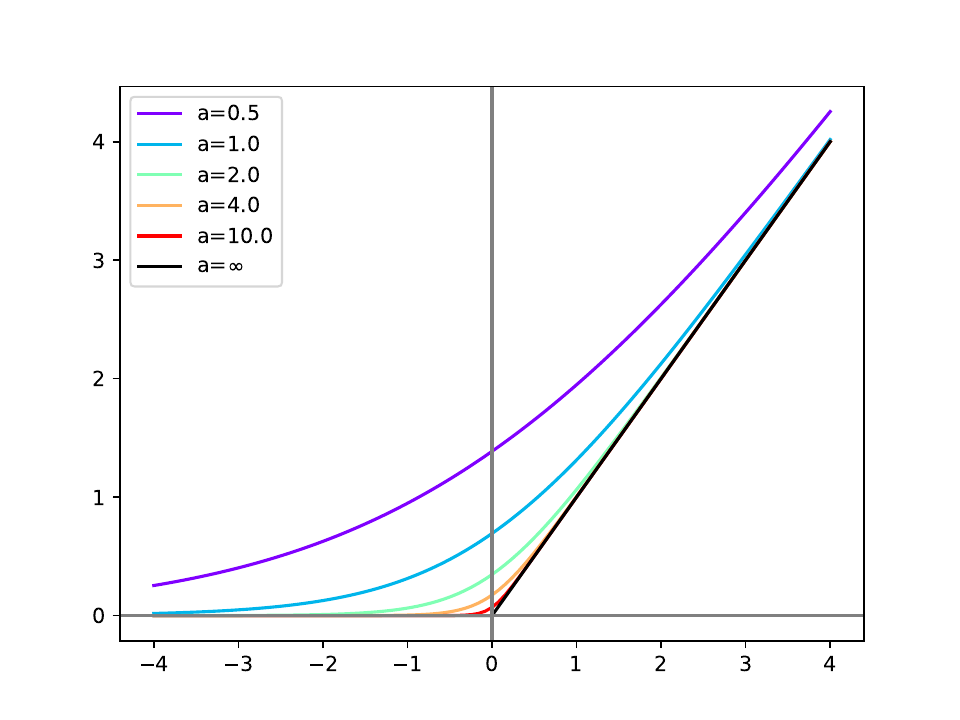}
\caption{Trainable rectifier $\operatorname{softplus}_a$ that smoothly approximates the ReLU ($a\to\infty$) and can represent the SoftPlus ($a=1$) exactly.}
\label{fig:softplus}
\end{figure}

For convention, we reference the free energy $\freeenergy$ to zero at the reference state (which is not at absolute zero temperature here).
The dissipation potential $\dissipationpotential$ is allowed to take on arbitrary reference values but has no flow (gradient) at  $\conjugateforce=\mathbf{0}$ and hence is just centered.
For elastic-plastic behavior, there must be a finite region around $\conjugateforce = \mathbf{0}$, as shown in \fref{fig:dissipation_potential}.
We penalize deviations from this assumption (more details are given in the next section), which also encourages the PANN not to fit mildly inelastic behavior as purely elastic.
We create this potential from a smooth potential $\varphi = \NN(\conjugateforce; \Zb)$ represented by a centered, partial ICNN, which we subsequently apply a rectifying function:
\begin{equation}
\phi(\conjugateforce,\Zc)     = \rect( \varphi(\conjugateforce,\Zc) )  \ge 0 \ .
\end{equation}
Its complement
\begin{equation}
\upsilon(\conjugateforce,\Zc) = \rect_c( \varphi(\conjugateforce,\Zc) ) = \varphi(\conjugateforce,\Zc) - \phi(\conjugateforce,\Zc) \le 0 \ ,
\end{equation}
acts as a yield-like indicator function.
Rate independence and the KKT conditions require the plastic flow state to remain at $\phi = 0$; however, we allow this to relax depending on the training data.
Hence, we employ an adaptive rectifier, illustrated in \fref{fig:softplus}:
\begin{equation}
\operatorname{softplus}_a (x) = a \log\left(1 + \exp\left(\frac{x}{a}\right) \right) \ ,
\end{equation}
with a trainable parameter $a$ such that  $\lim_{a\to 0} \operatorname{softplus}_a (x) = \relu(x)$ and $\operatorname{softplus}_1 (x) = \operatorname{softplus} (x)$.
This output activation also avoids issues with zero gradients and yet allows an arbitrary close approximation of the ReLU, which is equivalent to the Macaulay bracket.

\begin{figure}
\centering
\includegraphics[width=0.65\linewidth]{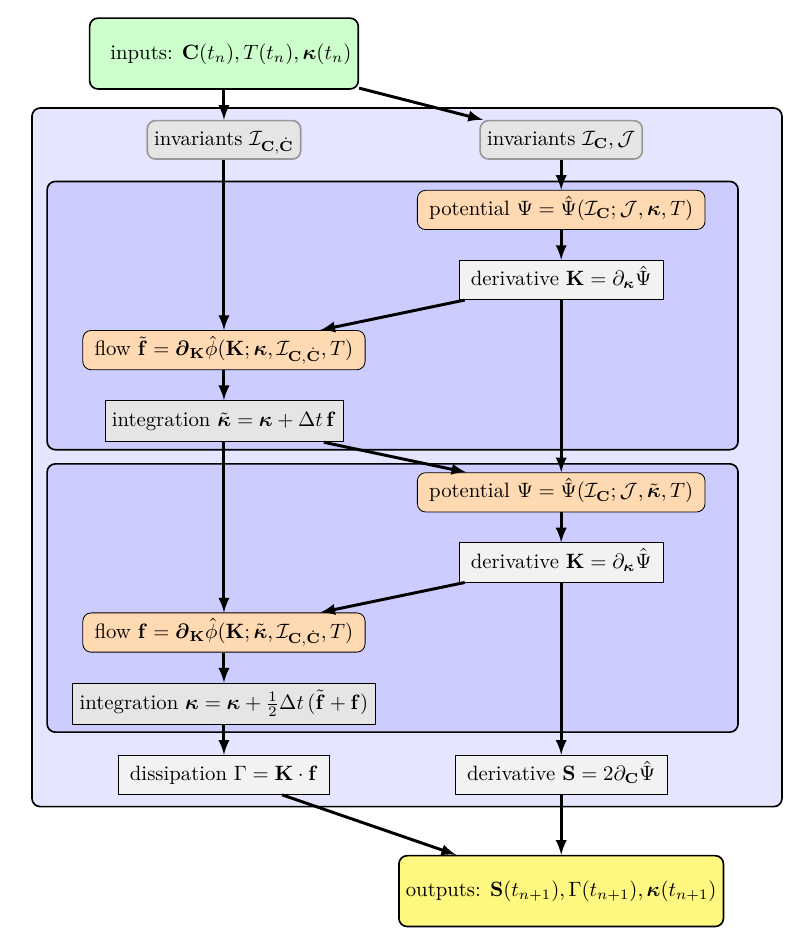}
\caption{The PANN framework for thermoviscoplasticity: Two potential ICNN-NODE integrated with a Heun-like scheme.}
\label{fig:heun}
\end{figure}

As in previous implementations of ISV neural networks \cite{jones2022neural,jones2025attention} we employ a neural ODE \cite{chen2018neural,dupont2019augmented} to integrate the flow
$\dot{\internalvariables} = \partialb_\conjugateforce \phi(\conjugateforce, \Zc)$
with a Heun-like integrator illustrated in \fref{fig:heun}.
We remark that recently other works have also employed neural networks to resolve the ODE system since it simplifies the training of the potentials, see \crefs{jones2022neural,rosenkranz2024viscoelasticty,holthusen2025complement}.

\subsection{Loss} \label{sec:loss}

\begin{figure}
\centering
\begin{subfigure}[c]{0.3\linewidth}
\includegraphics[width=0.99\linewidth]{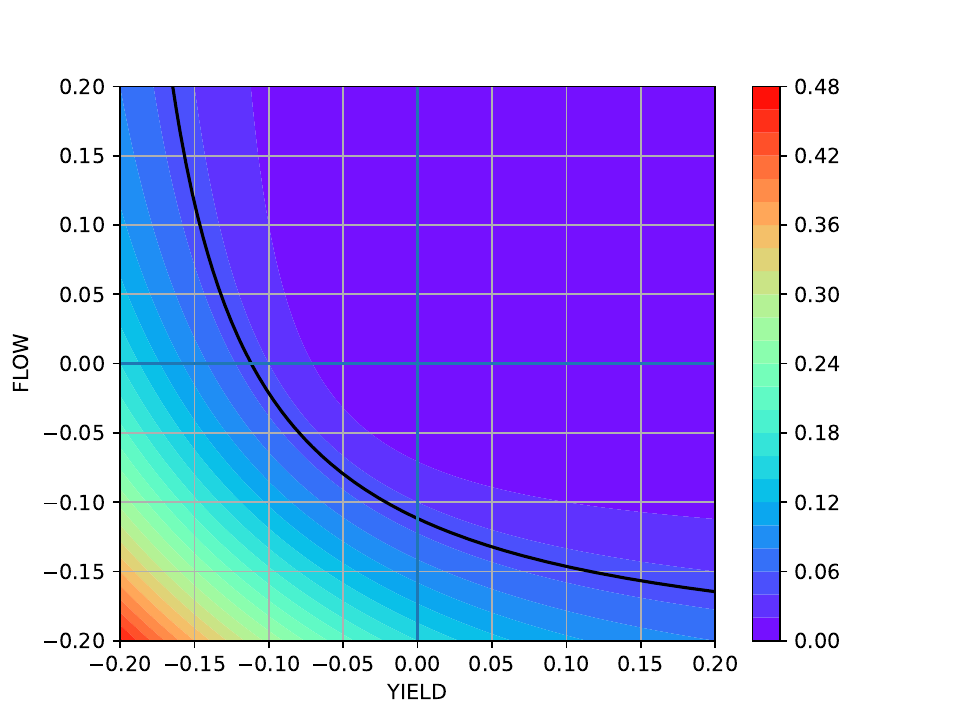}
\caption{$p=2$}
\end{subfigure}
\begin{subfigure}[c]{0.3\linewidth}
\includegraphics[width=0.99\linewidth]{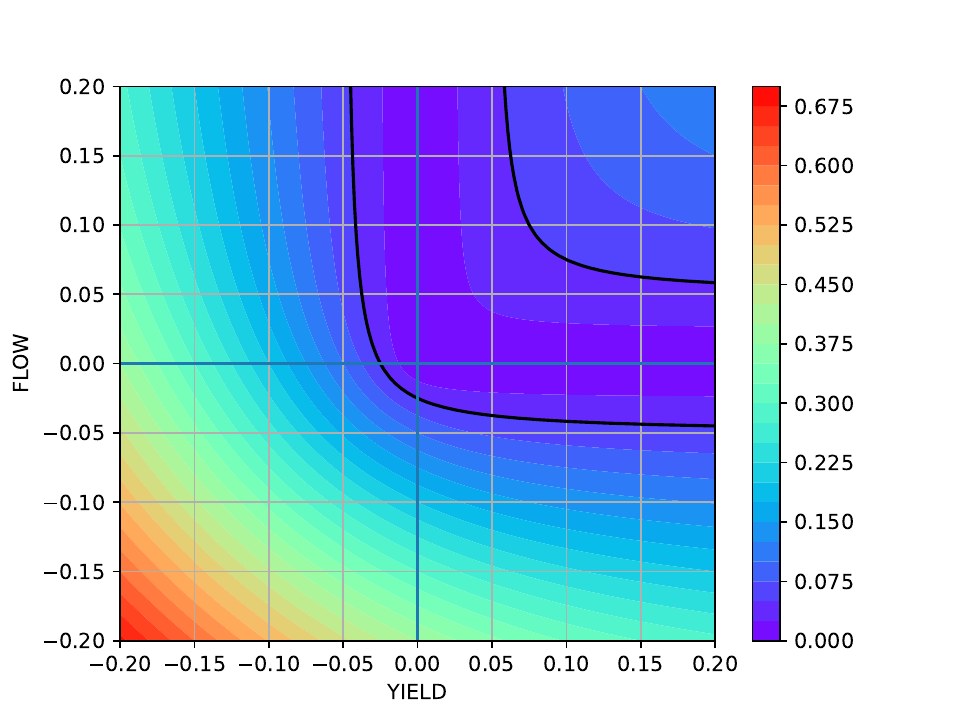}
\caption{$p=1$}
\end{subfigure}
\begin{subfigure}[c]{0.3\linewidth}
\includegraphics[width=0.99\linewidth]{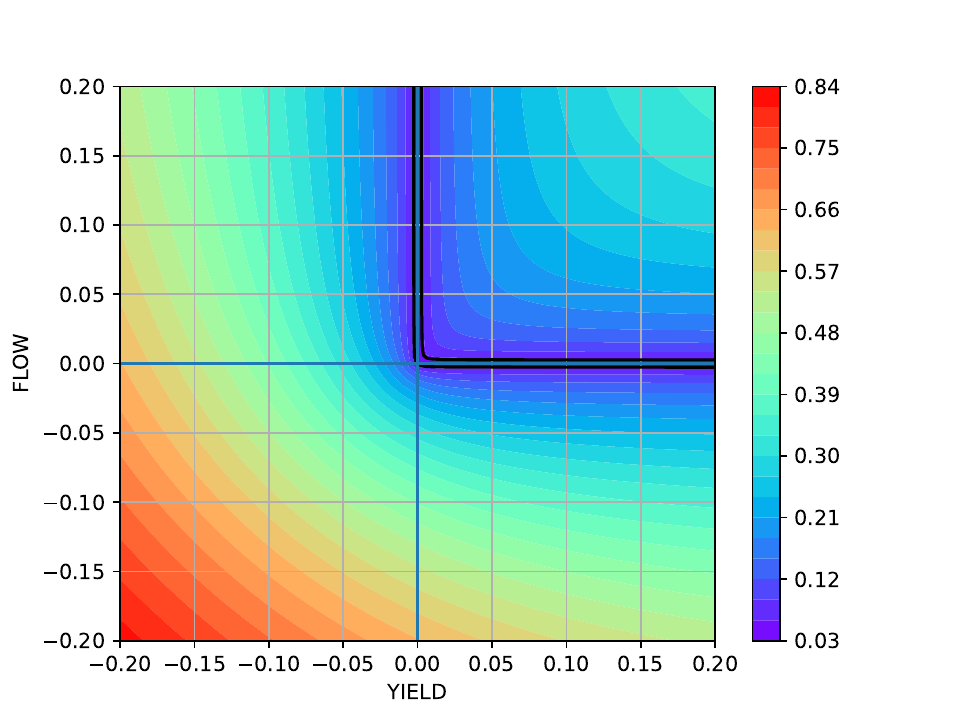}
\caption{$p=1/2$}
\end{subfigure}

\caption{Complementarity loss function diffused with p=2, nominal p=1, sharpened with power = 1/2
}
\label{fig:duality_loss}
\end{figure}

We take observable stress and temperature as functions of deformation trajectories as training data: $\data = \{ \Cb(t); \abstemperature(t), \pkstress(t) \}$.
The calibration loss:
\begin{equation} \label{eq:loss}
\Lc = \| \pkstress-\hat{\pkstress}\|^2_2 + \beta \| \abstemperature-\hat{\abstemperature}\|^2_2 + \lambda_\text{KKT} | \operatorname{ncf}_p(\upsilon,\| \dot{\internalvariables} \| ) |  + \lambda_e \| \langle \varphi(\mathbf{0} \rangle  \|^2_2 \ ,
\end{equation}
is a combination of:
(a) mean squared errors (MSE) of the stress and temperature using the  2-norm,
(b) a nonlinear complementary function (ncp)
encouraging duality between flow and a yield-like indicator
and
(c) the 2-norm of the Macaulay bracket of the potential $\varphi$ at the origin $\conjugateforce=\mathbf{0}$ to
enforce the existence of an elastic/no flow region.
Here,
$\beta$ is a dimensional scaling factor while the other penalty parameters ($\lambda_\text{KKT}$ and $\lambda_e$) act as a degree of belief.
In particular, a higher $\lambda_\text{KKT}$ is consistent with the belief that the dissipation should not access the positive region of the dissipation potential, the data is rate-independent, and that the KKT conditions should be exact.
Furthermore, a higher  $\lambda_e$ is associated with a stronger belief that the material has an elastic region (and is not purely viscoplastic), although it only constrains the reference state directly.
This interpretation is in loose analogy with Bayesian priors \cite{padmanabha2024improving}.
In our work, this treatment of the KKT conditions sharpen the flow transitions like a gate \cite{jones2025attention}.
Here we use a (modified/generalized) Fischer–Burmeister \cite{chen2008family} nonlinear complementarity problem function:
\begin{equation}
\operatorname{ncf}_p(\upsilon,\vargamma)
=  | ( \upsilon + \vargamma - \| \upsilon,\vargamma \|_2 ) |^p  \ ,
\end{equation}
where $\upsilon = \rect_c(\varphi)$ is the complement of the dissipation potential and $\vargamma = \| \partialb_\conjugateforce \dissipationpotential \|$ is a norm on the flow $\dot{\internalvariables}$..
This function is illustrated in \fref{fig:duality_loss} for a few choices of the hyperparameter $p$.
In this work, we fix $p= \tfrac{1}{2}$, but a regularizing objective $\lambda_2 | p |$ could be added to the loss $\Lc$ to allow for training of $p$, albeit at the expense of another penalty parameter.

\section{Data} \label{sec:data}

We generate a variety of training data for the demonstrations.
First, we survey a spectrum of viscoplastic behavior, then we fit the model to stochastic volume element (SVE) data that has no closed-form expression.
For the viscoplastic exemplars, it is sufficient and expedient to explore the ability of the proposed PANN model to represent rate dependence at constant temperature.
Then, we explore temperature dependence and rate effects with SVEs governed by crystal plasticity.
Lastly, we demonstrate how the model can be integrated into experimental calibration with a synthetic tension experiment where stress and temperature are measured.

\subsection{Path sampling}
To calibrate a path-dependent model, a variety of loading paths are needed.
Random walks \cite{bonatti2022importance} have been used, but they are difficult to bound within feasible loading limits without modification, such as reflecting boundaries.
Here we use periodic functions with random frequencies and amplitudes:
\begin{equation} \label{eq:F_samples}
\defgrad(t) = \sum_i \Ab_i \circ f(\Omegab_i t) \ ,
\end{equation}
where the tensor amplitudes $\Ab_i$ and frequencies $\Omegab_i$ are drawn from uniform distributions and $\circ$ is the Hadamard/element-wise product.
For simplicity, we only use a single term in \eref{eq:F_samples}.
Furthermore, for an isotropic material, it is sufficient to sample $\defgrad = \diag({F_{11}(t),F_{22}(t),F_{33}(t)})$ so that $\Ab \circ \Omegab = \Ab \Omegab$.
Hence
\begin{equation}
\Cb(t) \equiv \defgrad^T(t) \defgrad(t) = \Ab^2 \circ f(\Omegab^2 t)
\quad \text{and} \quad
\dot{\Cb}(t) = \Ab^2 \Omegab^2 \circ f(\Omegab^2 t)
\end{equation}
Defining $\ab$ to be the diagonal of $\Ab^2$ and $\omegab$ to be the diagonal of $\Omegab^2$, the principal invariants can be expressed as :
\begin{eqnarray}
I_1 &=&     \sum_i a_i f(\omega_i t) \\
I_2 &=& I_3 \sum_i 1/(a_i f(\omega_i t)) \\
I_3 &=&    \prod_i a_i f(\omega_i t)
\end{eqnarray}
We utilize harmonics (sines) for training due to their sampling of a range of rates in one trajectory, and sawtooth functions for out-of-distribution testing since the single rates are simpler to interpret.

\begin{figure}
\centering
\begin{subfigure}[c]{0.65\linewidth}
\includegraphics[width=0.95\linewidth]{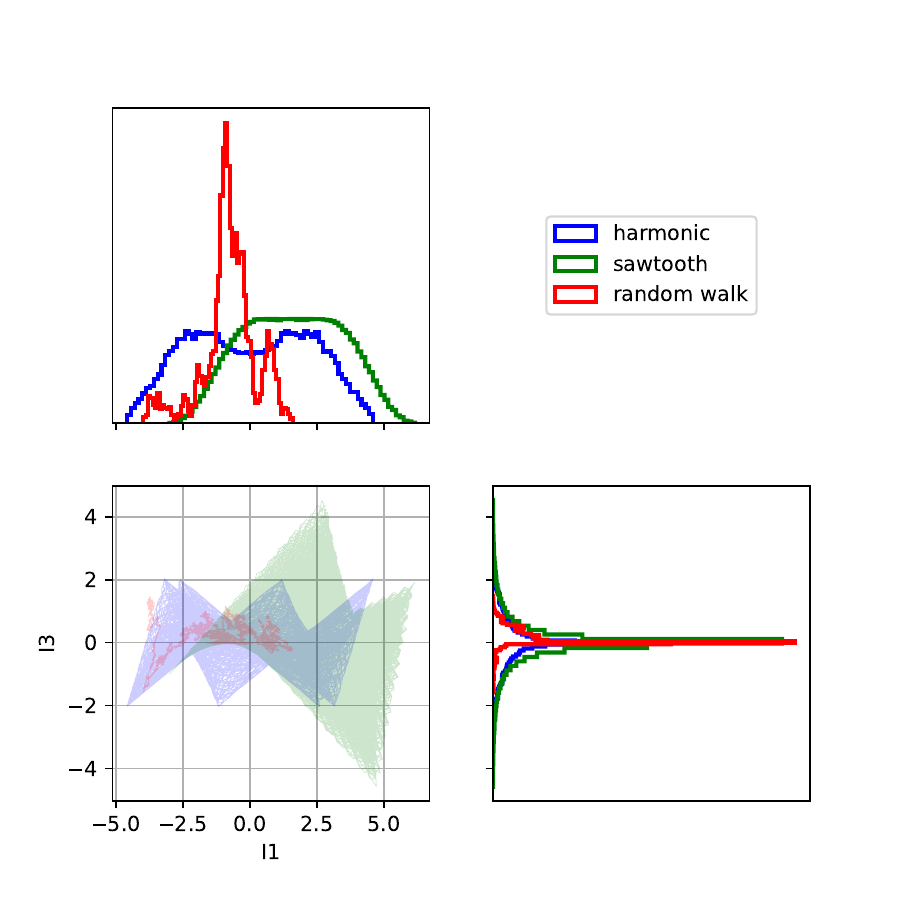}
\caption{deformation $I_1$ : deformation $I_3$}
\end{subfigure}
\begin{subfigure}[c]{0.65\linewidth}
\includegraphics[width=0.95\linewidth]{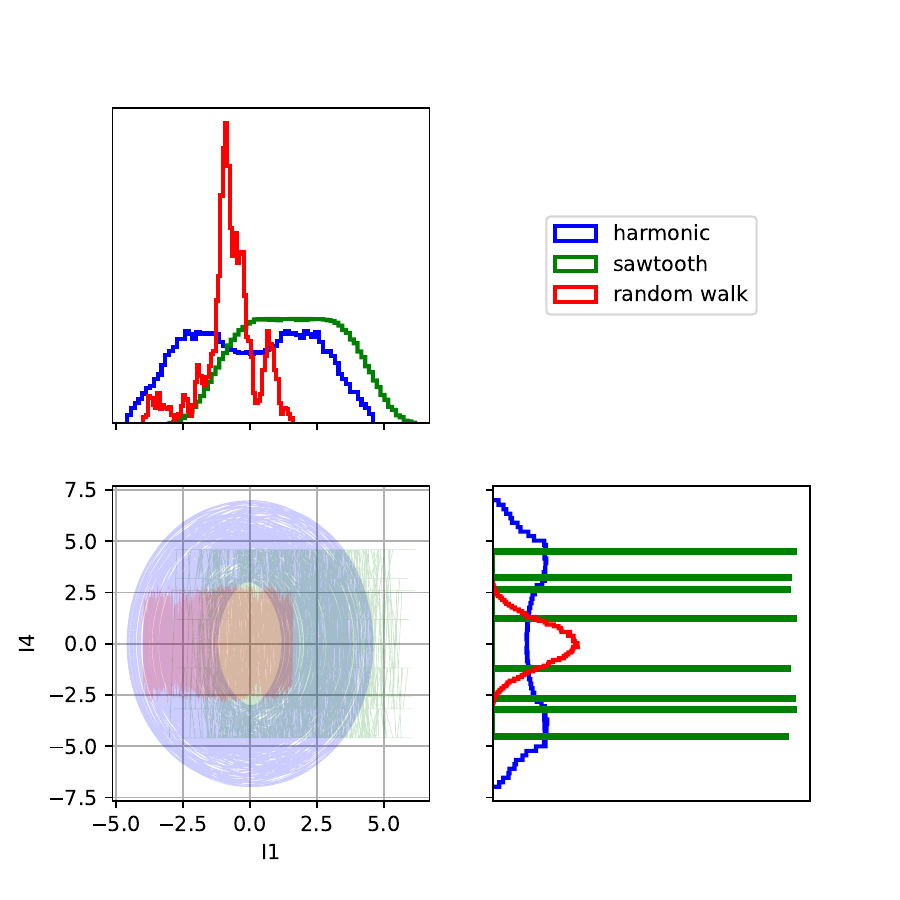}
\caption{deformation $I_1$ : rate $I_4$}
\end{subfigure}
\caption{Path sampling orbits: (a) deformation invariants $I_1 = \tr \Cb$ and $I_3 = \det \Cb$, and (b) deformation-rate invariants $I_1$ and $I_4 = \tr \Cb\dot{\Cb}$.
The lower left panels show the trajectories, while the upper and right panels depict the marginal densities.
}
\label{fig:sampling}
\end{figure}

\fref{fig:sampling} illustrates the deformation and rate invariant distributions for random walks, random harmonics, and random sawtooth trajectories.
Clearly, the harmonic ensemble does the best at uniformly sampling the invariants within a well-controlled range.
Note that the distribution of rates for the random walk is controlled by the distribution of displacement step sizes over a fixed time step.
Likewise, the velocity distribution of the sawtooth samples are Dirac deltas due to the constant rate stages of each path.%
\footnote{
For as single sine $x(t) = \sin(t)$, $\dot{x} = \cos(t)$, $dx = \cos(t) dt$, and $t = \arcsin(x)$,
$(\sqrt{1-x^2})^{-1} dx = dt$, so
\begin{equation*}
\operatorname{CDF}(x) \propto
\int_{-1}^{x} \frac{1}{| \dot{x}(t(y)) |} \, \mathrm{d}y
= \int_{-1}^{x} \frac{1}{\sqrt{1-y^2}} \mathrm{d}y
\end{equation*}
hence $\operatorname{PDF}(x) = \mathrm{d}_x \operatorname{CDF}(x) = \frac{1}{\pi \sqrt{A^2-x^2}}$ where $A$ is the amplitude, results from the Leibniz rule.
}

\subsection{Viscoplastic exemplars} \label{sec:viscoplastic_data}
As mentioned, we create data to cover 3 cases from the viscoplastic spectrum:
\begin{itemize}
\item \emph{Rate-independent elastoplasticity}:
has a finite elastic region, rate-independent flow, such as von Mises/J2 \cite{mises1913mechanik}, where the KKT constraints and multiplier are active.
\item \emph{Rate-dependent yield flow}:
has a finite elastic region, rate-dependent flow, such as the linear hardening model \cite{Simo1998} and the non-linear hardening Johnson-Cook \cite{johnson1983constitutive} model.
\item \emph{Pure viscoplasticity}:
has no elastic region,  all behavior is viscous, such that the internal state is always flowing and rate-dependent, such as the Perzyna model \cite{perzyna1971thermodynamic} (and Duvaut-Lions model \cite{duvaut1972inequations}) with a negligible yield strength.
\end{itemize}
We follow implementations in \cref{simo1992algorithms} and \cref{steinmann2021catalogue}.
Details of each are given in \aref{app:vp_exemplars}.

\begin{figure}
\centering
\begin{subfigure}[c]{0.24\textwidth}
\includegraphics[width=0.99\linewidth]{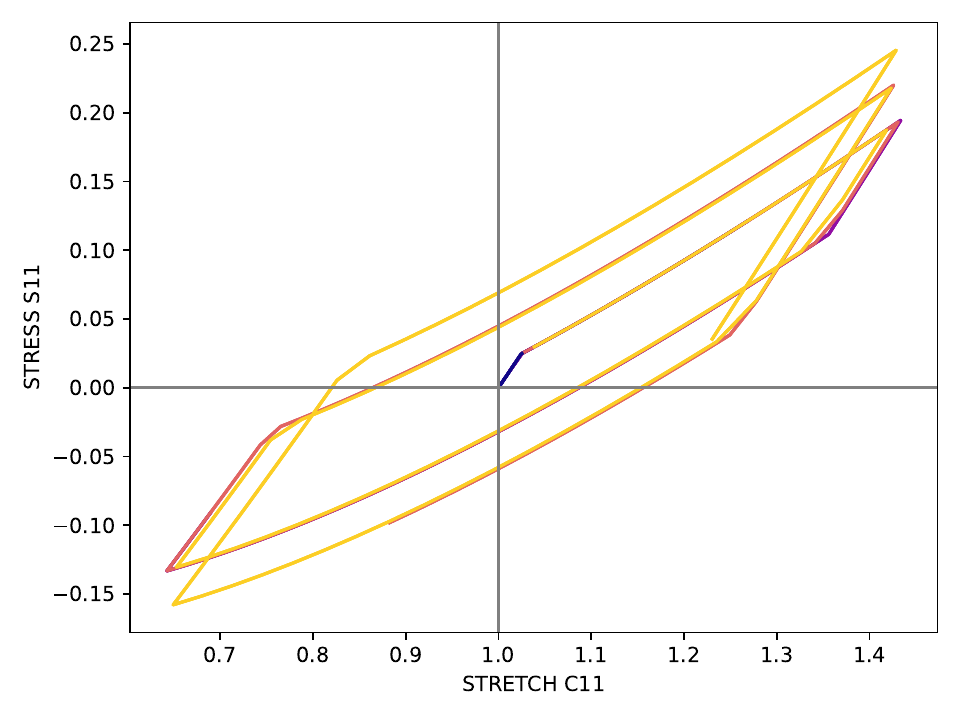}
\caption{rate-independent}
\end{subfigure}
\begin{subfigure}[c]{0.24\textwidth}
\includegraphics[width=0.99\linewidth]{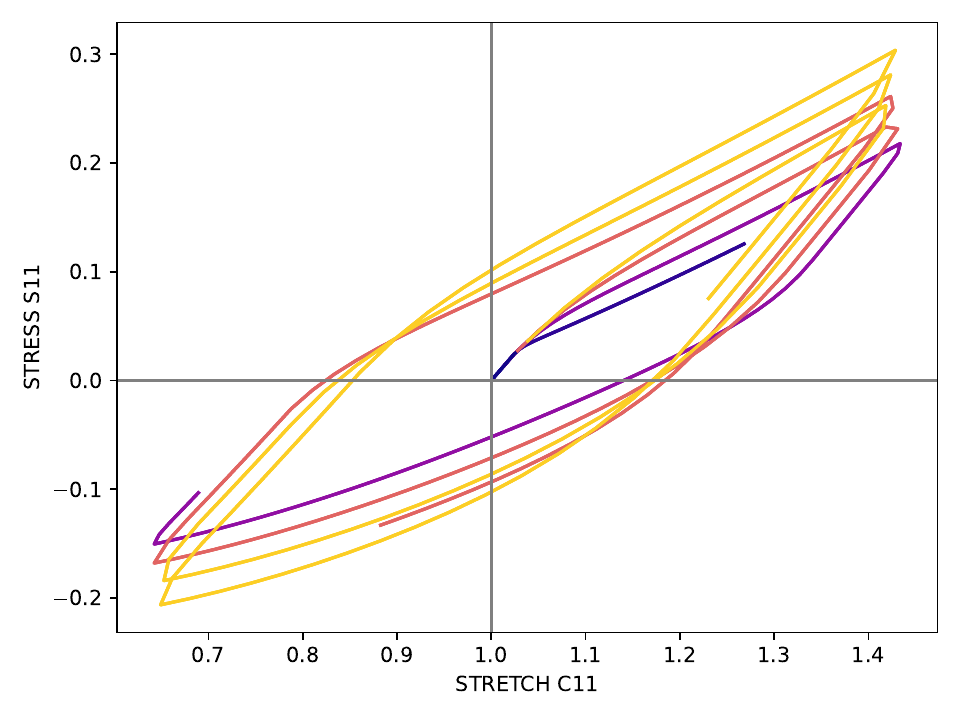}
\caption{Rate-dep. linear hardening}
\end{subfigure}
\begin{subfigure}[c]{0.24\textwidth}
\includegraphics[width=0.99\linewidth]{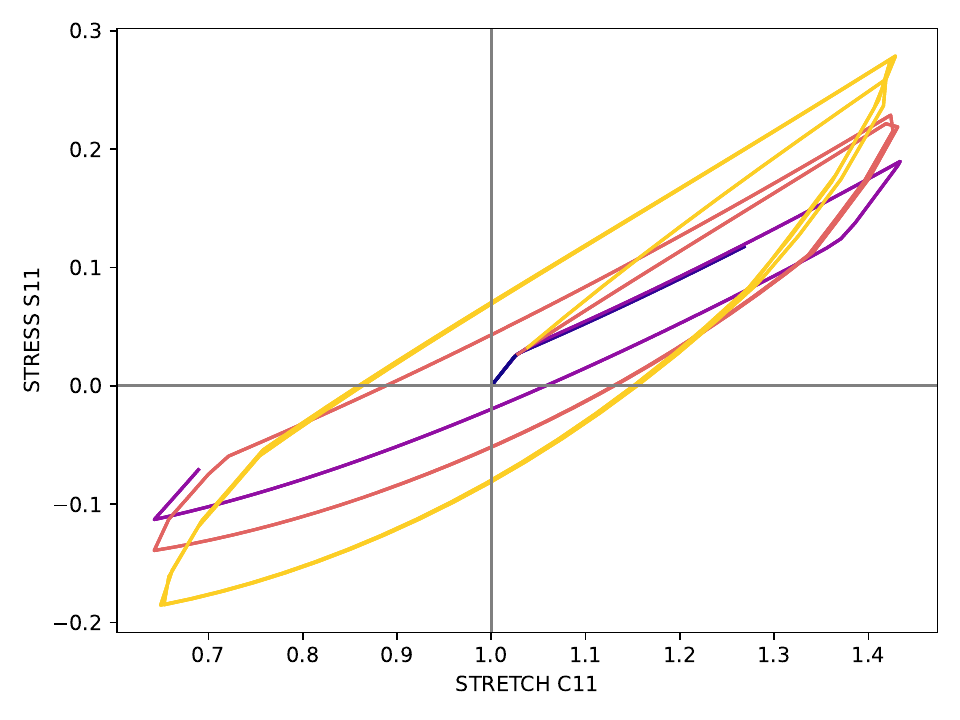}
\caption{Rate-dependent hardening}
\end{subfigure}
\begin{subfigure}[c]{0.24\textwidth}
\includegraphics[width=0.99\linewidth]{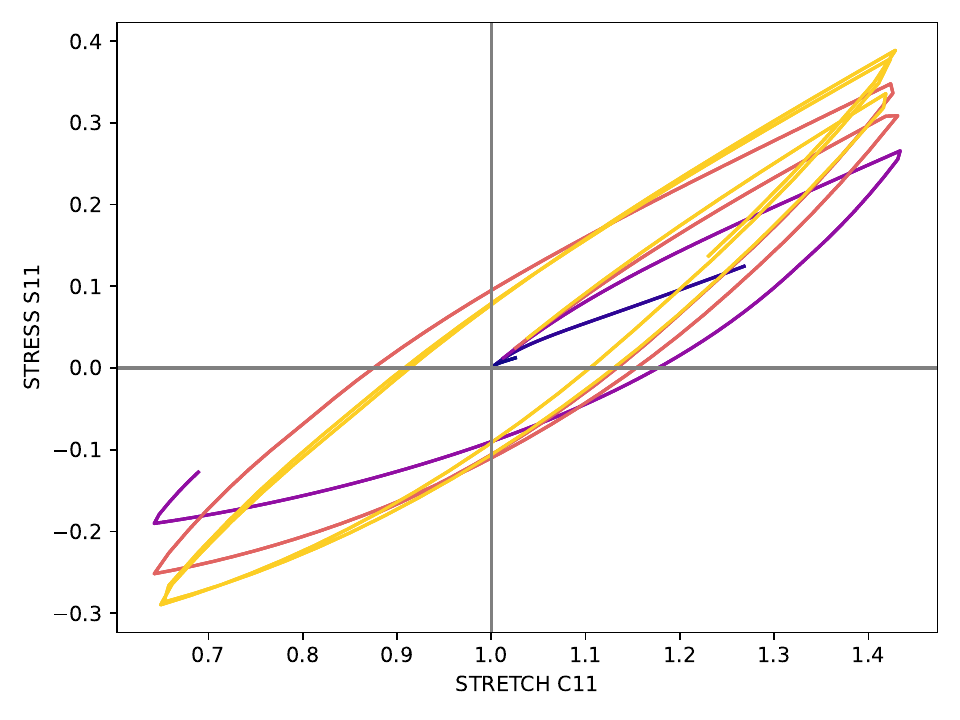}
\caption{Pure viscoplasticity}
\end{subfigure}
\caption{Exemplar (non-dimensionalized) viscoplastic data.
Colors indicate rates  (black:0.01, 0.1, 0.5, 1.0, 1.5:yellow).
}
\label{fig:exemplar_data}
\end{figure}

\fref{fig:exemplar_data} illustrates the phenomenological differences of the hysteresis behavior of the models.
\fref{fig:exemplar_data}a displays hardening but no rate dependence.
Both \fref{fig:exemplar_data}b,c show rate-dependent hardening but also have a finite elastic region, while \fref{fig:exemplar_data}d is characterized by rate dependence with a negligible elastic region.

\subsection{Microstructural  data}
\begin{figure}
\centering
\includegraphics[width=0.35\linewidth]{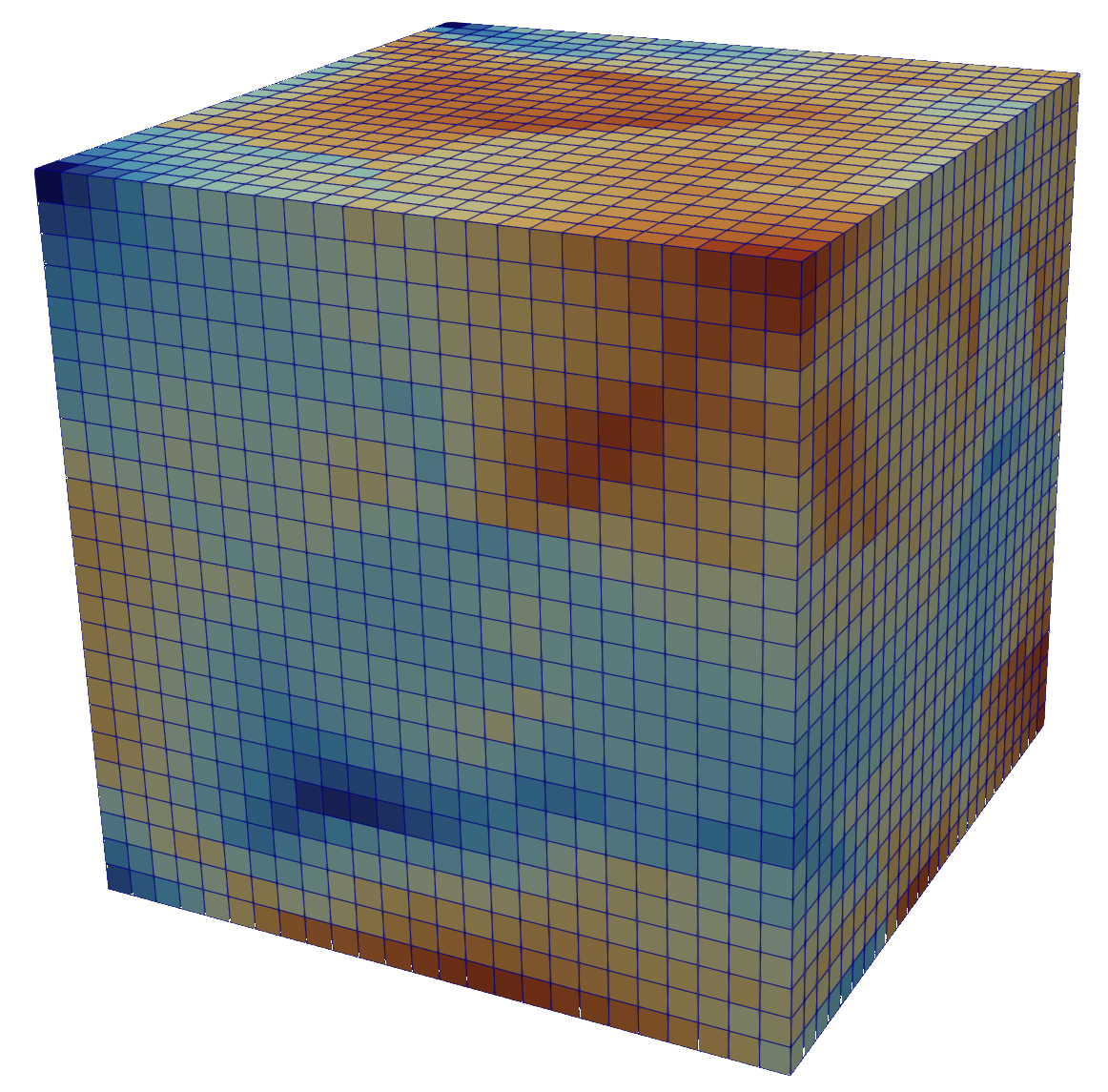}
\caption{A representative polycrystalline SVE (von Mises) stress field.}
\label{fig:cp_stress}
\end{figure}

\begin{figure}
\centering
\includegraphics[width=0.95\linewidth]{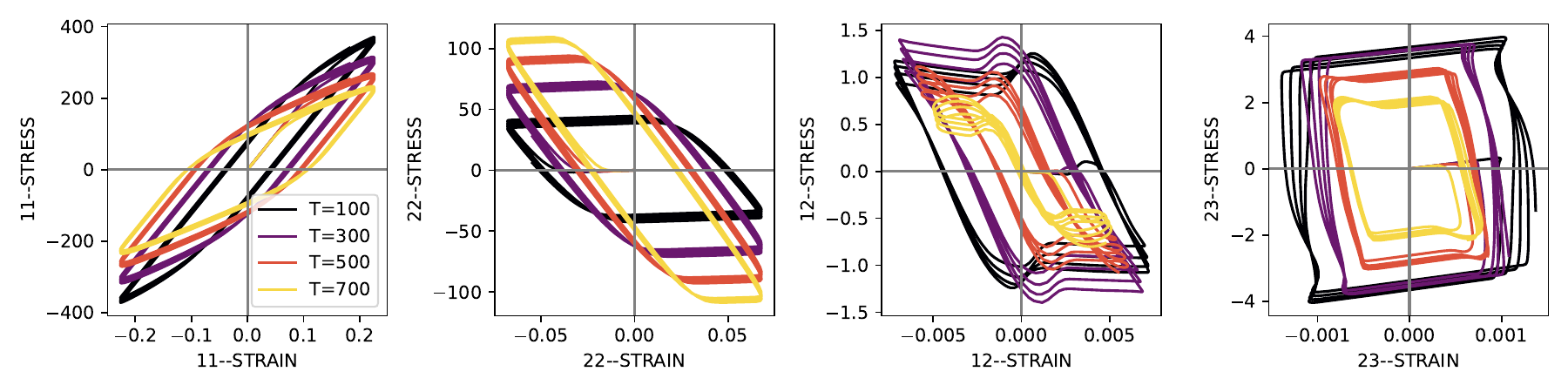}
\caption{Crystal plasticity: hysteresis response changes with temperature (black:100, 300, 500, 700 K:yellow).
}
\label{fig:cp_hysteresis}
\end{figure}

To demonstrate that the proposed framework performs well when the data comes from a response with no closed form and to show its utility as a closure in multiscale simulation, we generate stress-strain data from the deformation of polycrystalline aggregates at constant temperatures (due to limitations of the simulator).
The cubic stochastic volume elements (SVEs) are deformed using periodic boundary conditions, and the sample-averaged deformation (and rate) and stress are collected as training and testing data.
The constitutive equations determining the viscoplastic response of each of the crystals in the polycrystal are given in detail in \aref{app:sve}.
A realization of the polycrystalline aggregate represented on a 25$^3$ structured grid is shown in \fref{fig:cp_stress}; clearly, the stress field is distinctly heterogeneous.
\fref{fig:cp_hysteresis} shows the changes in hysteresis over a range of temperatures.

\subsection{Simulated tension-heating experiment}

For this demonstration, we emulate a high-rate tension experimental setup with heating due to plastic work of the specimen and ambient (Newton) cooling.

For the (synthetic) material response, we make the common assumption of a separable free energy \cite{ottosen2005mechanics} :
\begin{equation}
\freeenergy =  \freeenergy_e(\lagrangestrain_e,\temperature)
+ \freeenergy_p(\internalvariables,\temperature,...)
+ \freeenergy_{\abstemperature}(\temperature)
\end{equation}
with the exception of the thermoelastic coupling \cite{armero1992new})
\begin{equation}
\freeenergy_e  =
\frac{1}{2} \lagrangestrain_e \Cbb \lagrangestrain_e + \temperature \thermalexpansiontensor \lagrangestrain_e
=
\frac{1}{2} K \, \tr^2 \lagrangestrain_e
+  G \, \dev \lagrangestrain_e : \dev \lagrangestrain_e
+  \thermalexpansioncoefficient \temperature \tr \lagrangestrain_e \, .
\end{equation}
We have introduced the empirical temperature $\temperature = \abstemperature - \abstemperature_0$, where $\abstemperature_0$ is an (absolute) reference temperature at which there is no thermal strain,
i.e.,  $\pkstress(\lagrangestrain=\mathbf{0},\abstemperature=\abstemperature_0) = \mathbf{0}$.
Here, $K$ is the bulk modulus, $G$ is the shear modulus, and $\beta$ is the thermal expansion.
We follow \cref{ottosen2005mechanics} and choose
\begin{equation}
\freeenergy_\abstemperature = \heatcapacity_\strain \left(\abstemperature - \abstemperature \log\left(\frac{\abstemperature}{\abstemperature_0} \right) \right) \ ,
\end{equation}
which is concave,
and
\begin{equation}
\freeenergy_p =  Y_\infty \eqps + (Y_\infty - Y_0) \left( 1 - \epsilon_0 \exp\left(-\frac{\eqps}{\epsilon_0}\right) \right) \ , 
\end{equation}
which corresponds to a rate-dependent Voce \cite{voce1948relationship} relation, i.e., $h = (Y_\infty + (Y_\infty - Y_0)(1- \exp(-\tfrac{\eqps}{\epsilon_0}) )$.
Rate dependence was modeled with an overstress relaxation with time constant $\tau$.
Hence, the $\freeenergy$ is consistent with the following constitutive models
\begin{alignat}{3}
\pkstress &\equiv& \density_0 \partialb_{\lagrangestrain} \freeenergy
&= ( K \tr \lagrangestrain_e + \thermalexpansioncoefficient \temperature ) \Ib  + 2 G \dev \lagrangestrain_e\\
\entropy &\equiv& - \density_0 \partial_\abstemperature \freeenergy
&=  \heatcapacity_\strain \log\left(\frac{\abstemperature}{\abstemperature_0}\right) - \thermalexpansioncoefficient  \tr \lagrangestrain_p
\\
\internalenergy &\equiv& - \abstemperature^2 \partial_\abstemperature \left(\frac{\freeenergy}{\abstemperature}\right)
&= \heatcapacity_\strain \abstemperature
+ \frac{1}{2} K \, \tr^2 \lagrangestrain_e
+  G \, \dev \lagrangestrain_e : \dev \lagrangestrain_e \\
\heatcapacity &\equiv& - \abstemperature \partial^2_\abstemperature \density \freeenergy
&= \heatcapacity_\strain
\end{alignat}
which is a version of the St. Venant model \cite{horgan2003finite}.
Note that the thermal expansion does not appear in the internal energy $\internalenergy$; and the temperature dependence of $\freeenergy_p$ complicates the interpretation of the heat capacity, hence we omit this dependence.

For this choice \cite{rosakis2000thermodynamic}, the heat equation \eref{eq:heat_eqn} becomes:
\begin{equation} \label{eq:tension_heat_equation}
\density \heatcapacity_\strain \dot{\temperature}
=
\overbrace{
\temperature \thermalexpansiontensor \dot{\lagrangestrain}_e
+
\underbrace{
\pkstress : \dot{\lagrangestrain}_p
- h \dot{\alpha}
}_{\dissipation_p}
}^{Q_p}
- \grad\cdot\heatflux +  \underbrace{\density \heatgeneration}_{ \zeta \temperature }
\end{equation}
where plastic work is a heat source, while hardening is a heat sink.
Here $\zeta \temperature$ is Newton cooling with coefficient $\zeta$.
Likewise, the dissipation inequality reduces to :
\begin{eqnarray} \label{eq:dissipation_reduced}
-\density (\dot{\freeenergy} + \entropy \dot{\abstemperature})
+ \pkstress : \dot{\lagrangestrain}
- \frac{1}{\abstemperature} \heatflux \cdot \grad \abstemperature
&=&   \\
\underbrace{(\pkstress -\density \partial_\lagrangestrain {\freeenergy})}_{=0} \dot{\lagrangestrain}
+ \underbrace{
\pkstress : \dot{\lagrangestrain}_p
+ \density \partial_{\internalvariable_a} \freeenergy \, \dot{\internalvariable}_a
}_{\dissipation_p}
-
\underbrace{\frac{1}{\abstemperature} \heatflux \cdot \grad \abstemperature
}_{\dissipation_\abstemperature}
\nonumber
&\ge& 0
\end{eqnarray}
which can be split into mechanical $\dissipation_\strain$ and thermal dissipation $\dissipation_\abstemperature$.
The usual Fourier conduction model
\begin{equation}
\heatflux = - \thermalconductivity \grad \temperature
\end{equation}
completes the system.
Finally, the Taylor-Quinney ratio of heating to plastic work \cite{rosakis2000thermodynamic,simo1992associative} is:
\begin{equation}
q = \frac{Q_p}{\Gamma_p} =
\frac{\temperature \thermalexpansiontensor : \dot{\lagrangestrain}_e
+ \pkstress : \dot{\lagrangestrain}_p
- h \dot{\alpha}}
{\pkstress : \dot{\lagrangestrain}_p} \, .
\end{equation}

\section{Results} \label{sec:results}

In this section, we present the results of training the proposed model on the three demonstrations whose data are described in the previous section.
We use the viscoplastic-spectrum exemplars to explore the basic properties of the PANN, including visualizing the potentials and investigating the effects of the penalty parameters (described in \sref{sec:loss}).)
The stochastic volume homogenization and synthetic tension calibration exemplars illustrate the utility of the framework in these common applications.

We were able to obtain good results on each of the three demonstrations with the same PANN architecture.
It had 9 internal states, 3 stretch and 7 stretch-rate invariants, and 3 layers each in the free energy (524 parameters) and dissipation potential (1675 parameters).
All activation functions were smooth softplus activations.

\subsection{Range of viscous behavior}
In this exemplar, we demonstrate that the proposed PANN can represent a full range of viscoplastic behavior, as described in \sref{sec:viscoplastic_data}, without architecture or training modifications.
In fact, all architectural and training parameters were the same for each trial, e.g., we used $\lambda_e = 10^{-4}$ and $\lambda_\text{KKT} = 10^{-8}$.
Here we use datasets of 1000 sinusoidal trajectories of each phenomenological model described in \sref{sec:viscoplastic_data}.
We split the data 70/10/20 into training, in-training validation, and held-out testing, respectively.
We also used 200 sawtooth trajectories for out-of-distribution testing.

\begin{figure}
\centering
\begin{subfigure}[c]{0.79\textwidth}
\centering
rate-dependent hardening \\
\includegraphics[width=0.99\linewidth]{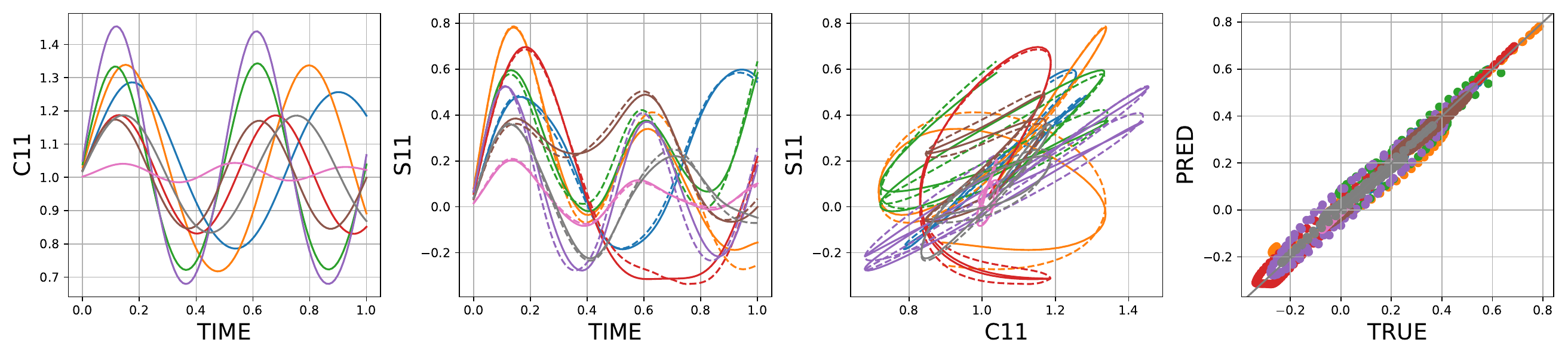} \\
viscoplasticity  \\
\includegraphics[width=0.99\linewidth]{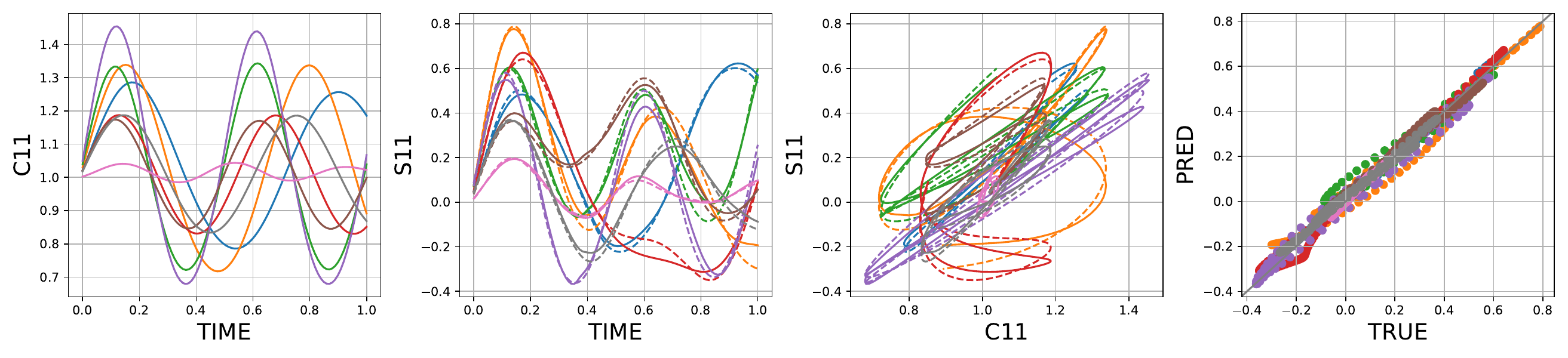}
\caption{in-distribution}
\end{subfigure}
\begin{subfigure}[c]{0.79\textwidth}
\centering
rate-dependent hardening \\
\includegraphics[width=0.99\linewidth]{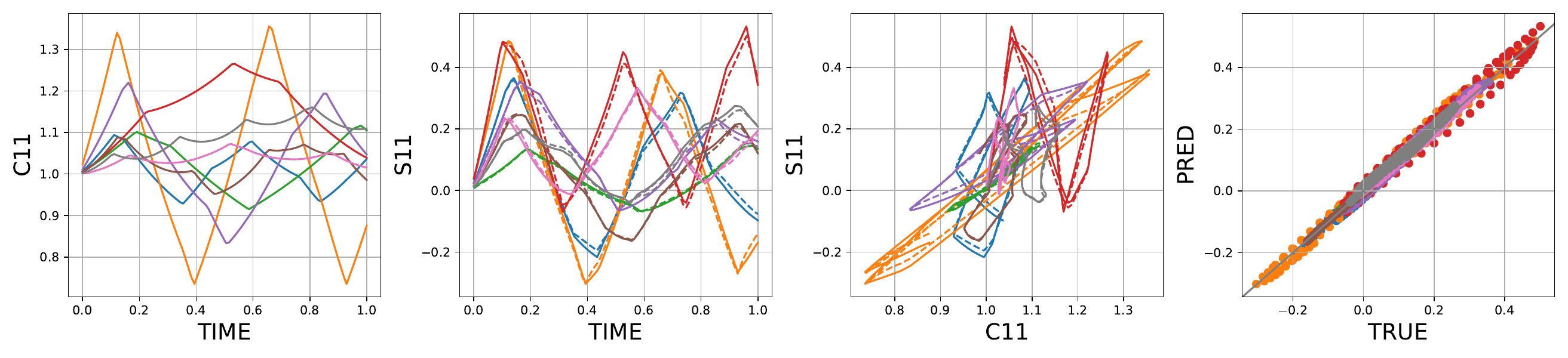} \\
viscoplasticity  \\
\includegraphics[width=0.99\linewidth]{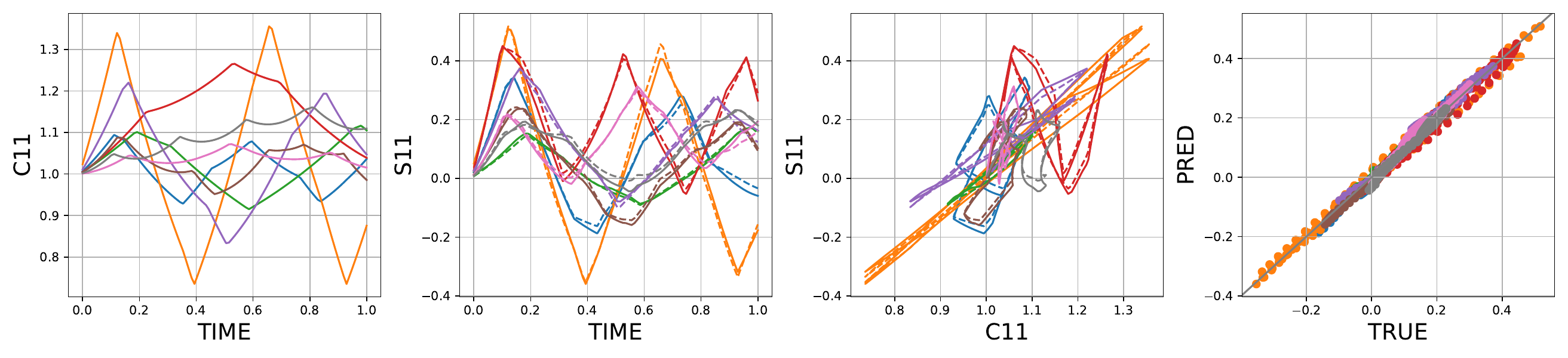}
\caption{out-of-distribution}
\end{subfigure}
\caption{Viscoplastic exemplars: comparison of predictions (solid lines) with held out data (dashed lines).
Note the results for the rate-independent and rate-dependent linear hardening cases are similar to those show.
The RMSE scores are summarized in \tref{tab:viscoplastic_errors}.
}
\label{fig:viscoplastic_results}
\end{figure}

\fref{fig:viscoplastic_results} illustrates that the model can accurately represent inelastic behavior ranging from rate-independent plasticity through rate-dependent hardening to fully viscous behavior.
There are small discrepancies between the predictions of the model and the held out data, but the model has excellent correlations and does comparably well in predicting both in-distribution trajectories (sine loading) and out-of-distribution trajectories (sawtooth loading), as shown in \fref{fig:viscoplastic_results} and summarized in \tref{tab:viscoplastic_errors}.
In fact, the 2--3\%  errors for a fairly small PANN are comparable across the various inelastic phenomenology and in- versus out-of-distribution held-out predictions.

\begin{figure}
\centering
\begin{subfigure}[c]{0.3\linewidth}
\includegraphics[width=0.9\linewidth]{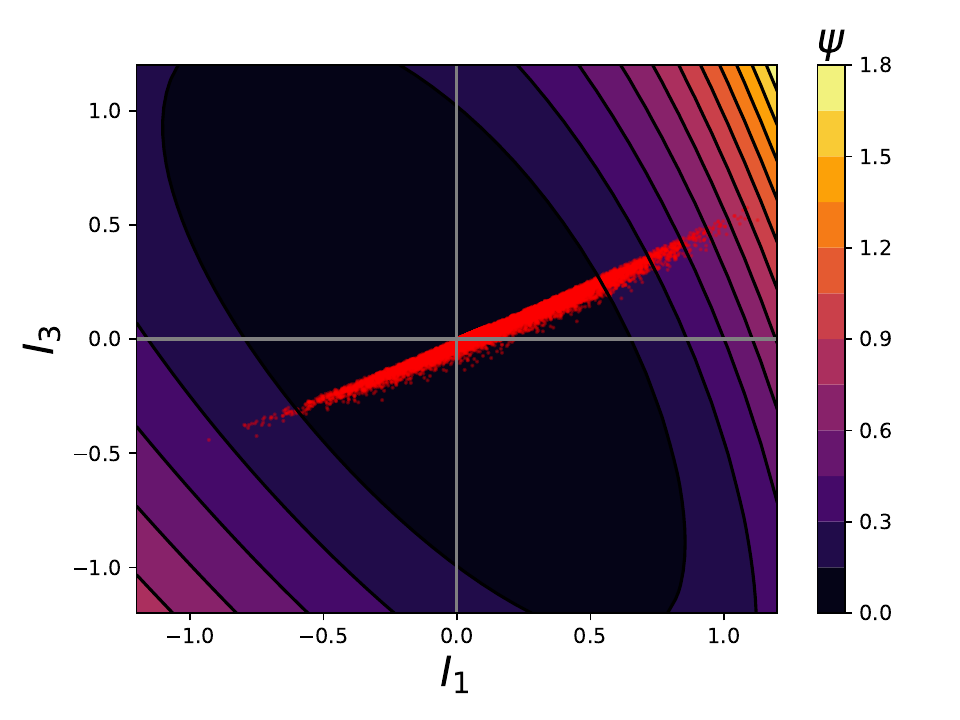}
\caption{invariant $I_1$ : $I_3$}
\end{subfigure}
\begin{subfigure}[c]{0.3\linewidth}
\includegraphics[width=0.9\linewidth]{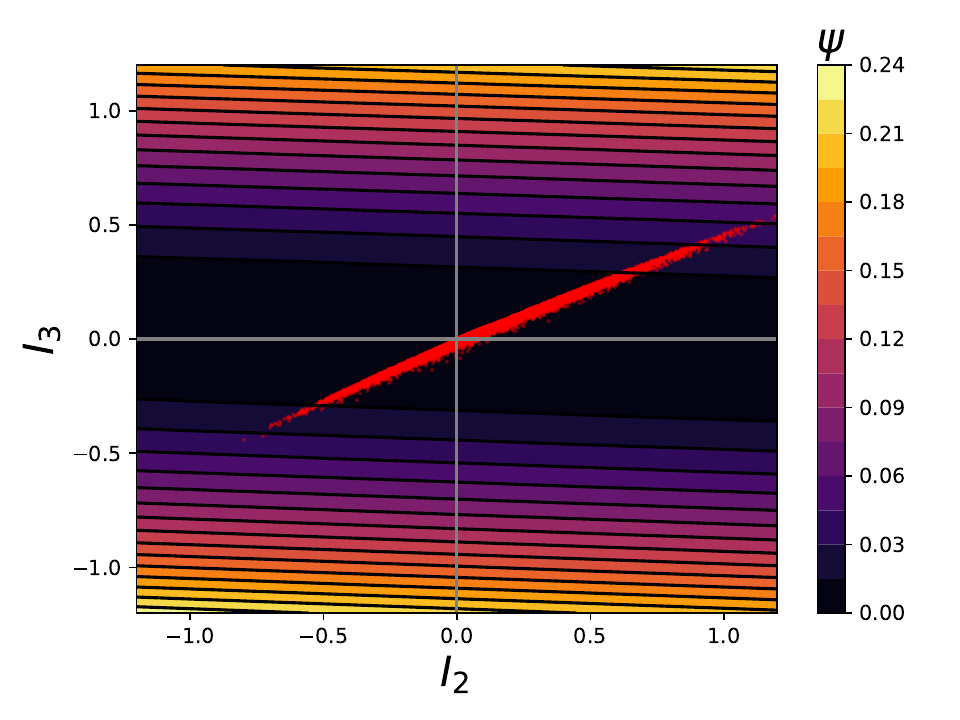}
\caption{invariant $I_2$ : $I_3$}
\end{subfigure}
\begin{subfigure}[c]{0.3\linewidth}
\includegraphics[width=0.9\linewidth]{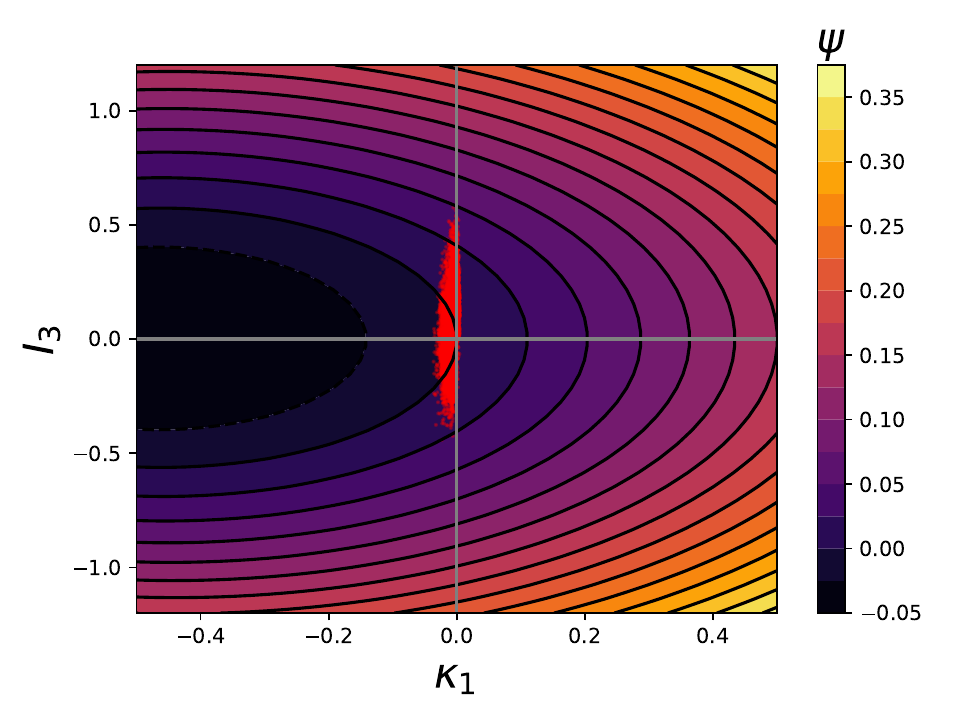}
\caption{internal state $\kappa_1$ : invariant $I_3$}
\end{subfigure}

\begin{subfigure}[c]{0.3\linewidth}
\includegraphics[width=0.9\linewidth]{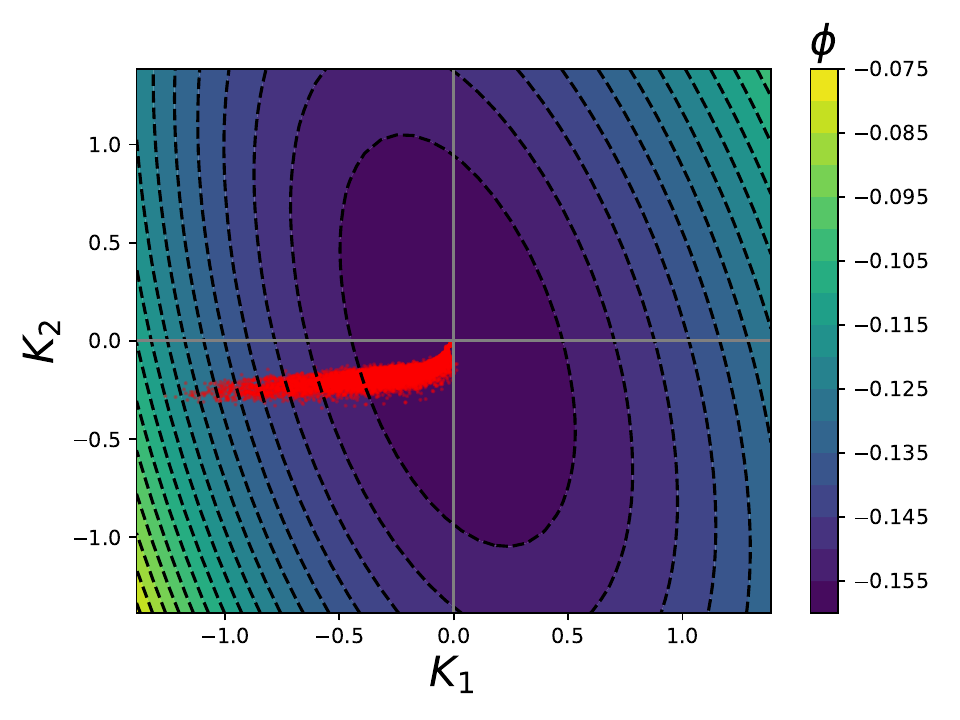}
\caption{conjugate force $K_1$ : $K_2$}
\end{subfigure}
\begin{subfigure}[c]{0.3\linewidth}
\includegraphics[width=0.9\linewidth]{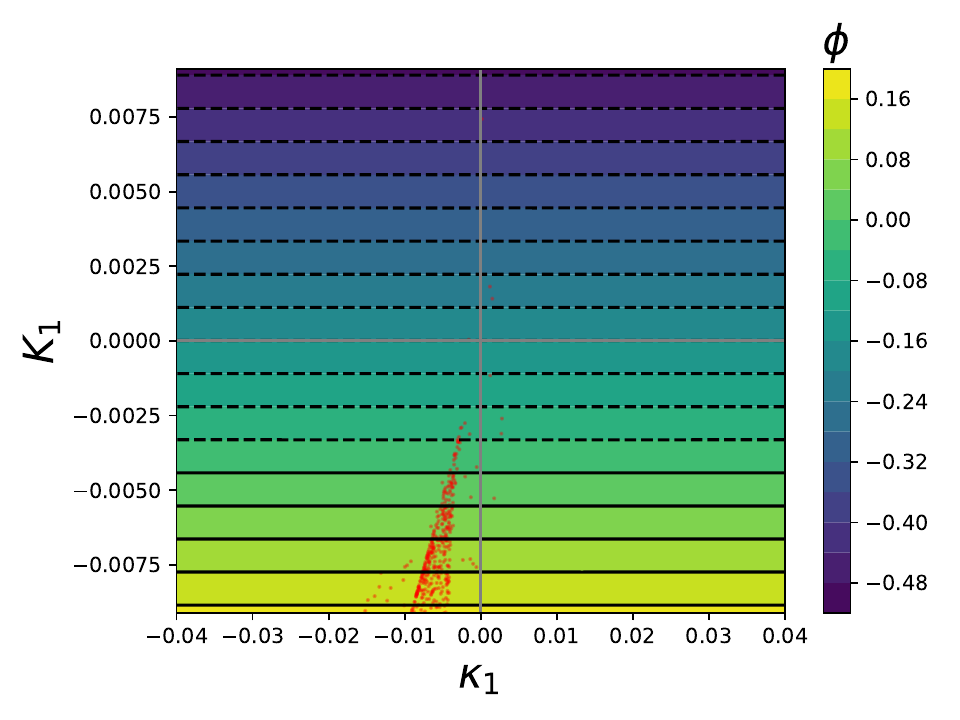}
\caption{state $\kappa_1$ : conjugate force $K_1$}
\end{subfigure}
\begin{subfigure}[c]{0.3\linewidth}
\includegraphics[width=0.9\linewidth]{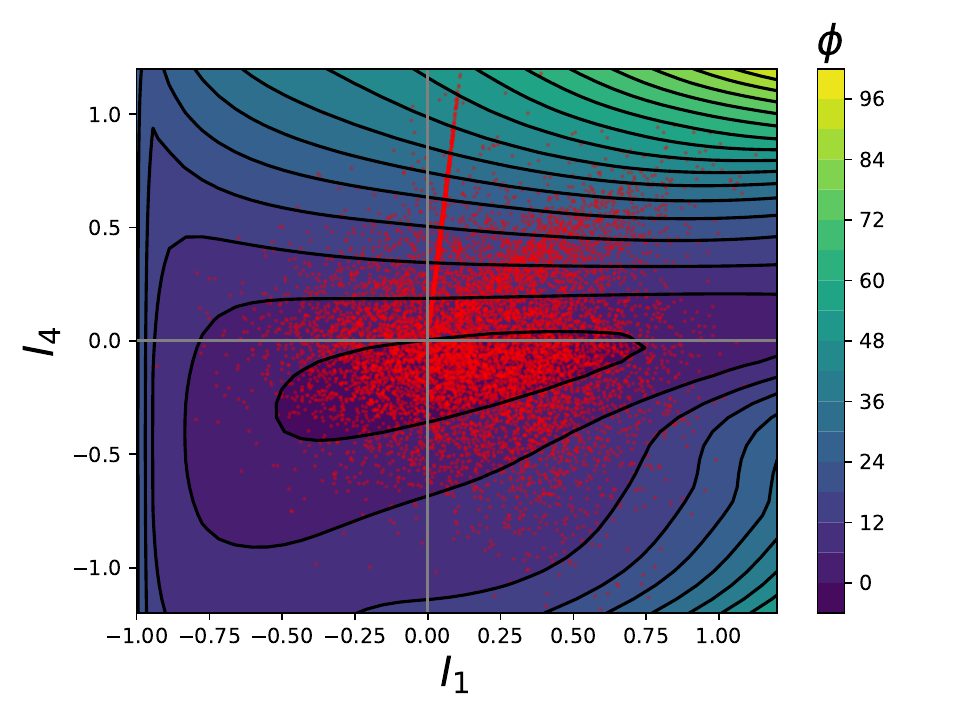}
\caption{invariant $I_1$ : $I_4$.}
\end{subfigure}

\caption{Visualization of the free energy $\psi$ and dissipation $\phi$ potentials for the rate-independent hardening data.
Red points are samples from the prediction trajectories.
Free energy
(a) invariant $I_1$ : invariant $I_3$,
(b) invariant $I_2$ : invariant $I_3$,
(c) internal state $\kappa_1$ : invariant $I_3$.
Dissipation potential
(d) conjugate force $K_1$ : conjugate force $K_2$,
(e) internal state $\kappa_1$ : conjugate force $K_1$,
(f) invariant $I_1$ : invariant $I_4$.
}
\label{fig:viz_j2}
\end{figure}

\begin{figure}
\centering
\begin{subfigure}[c]{0.3\linewidth}
\includegraphics[width=0.9\linewidth]{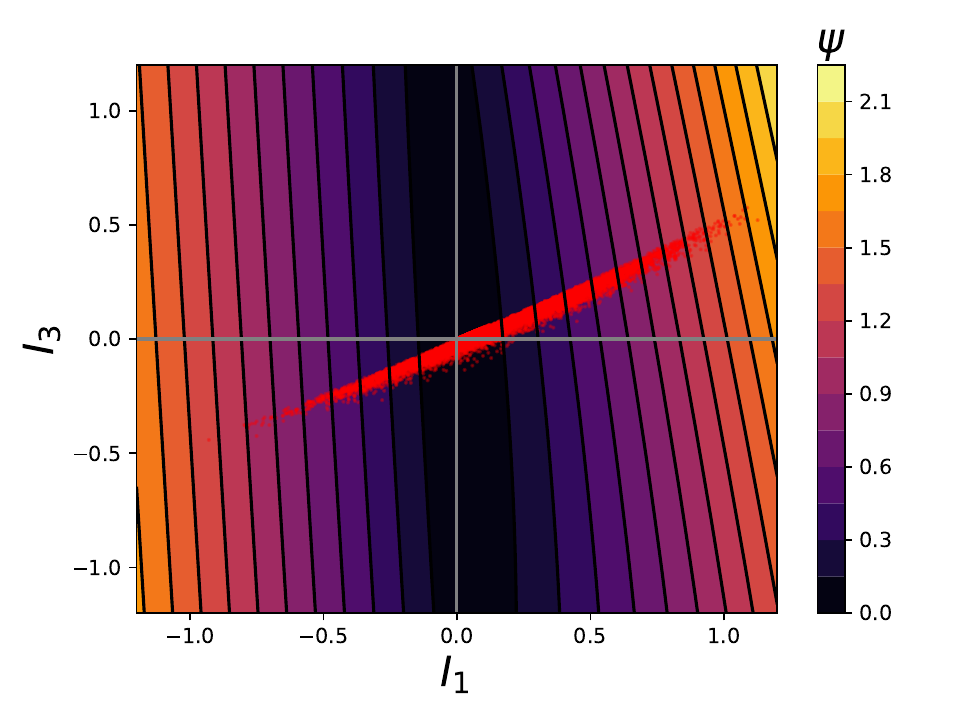}
\caption{invariant $I_1$ : $I_3$}
\end{subfigure}
\begin{subfigure}[c]{0.3\linewidth}
\includegraphics[width=0.9\linewidth]{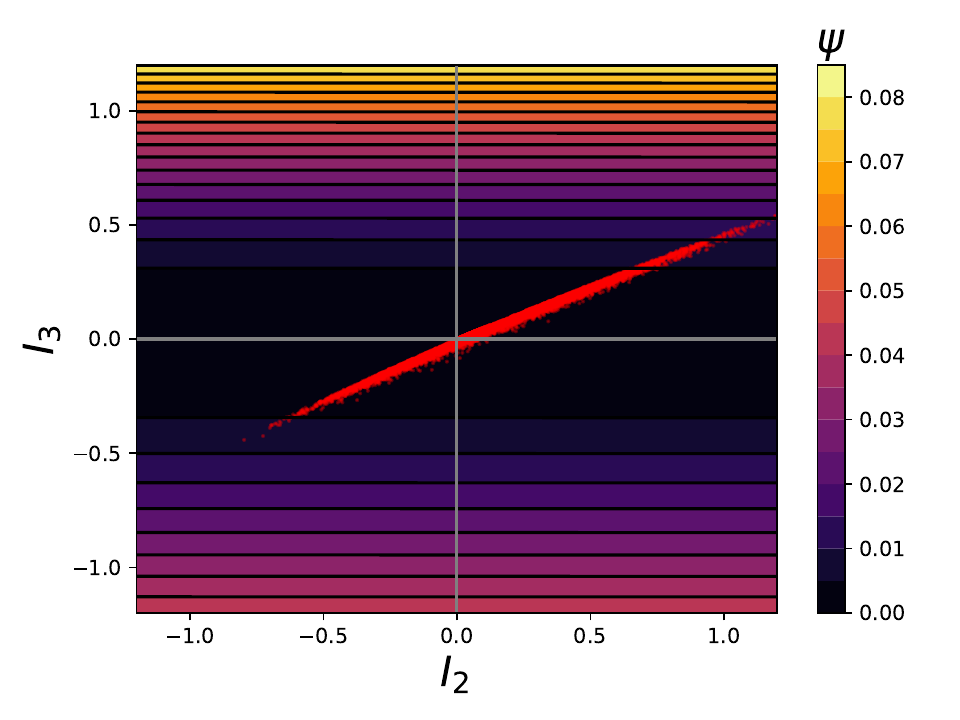}
\caption{invariant $I_2$ : $I_3$}
\end{subfigure}
\begin{subfigure}[c]{0.3\linewidth}
\includegraphics[width=0.9\linewidth]{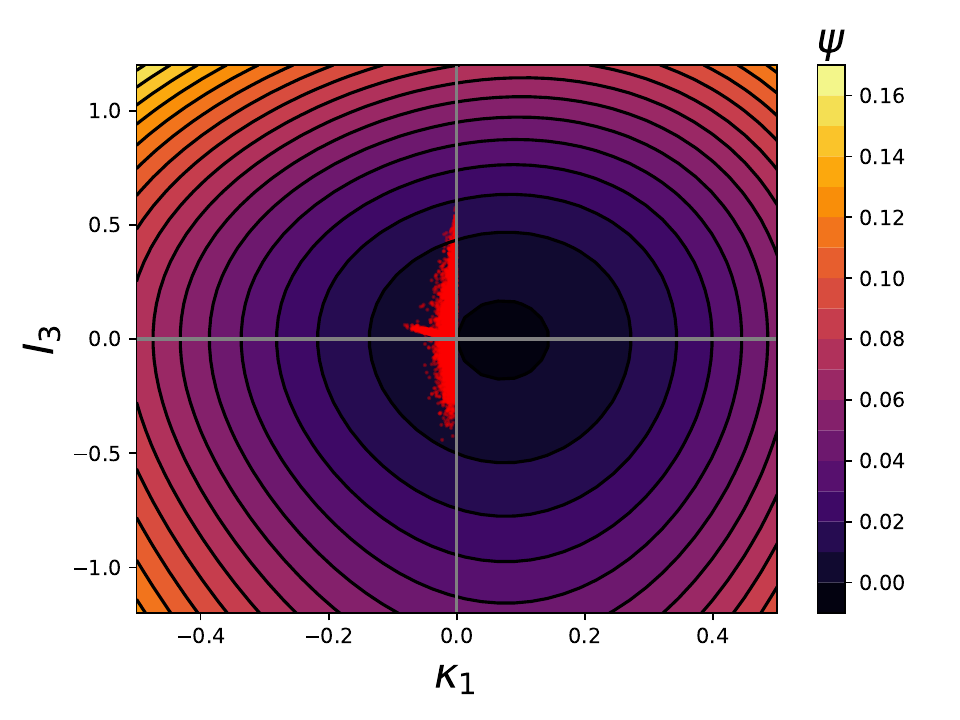}
\caption{internal state $\kappa_1$ : invariant $I_3$}
\end{subfigure}

\begin{subfigure}[c]{0.3\linewidth}
\includegraphics[width=0.9\linewidth]{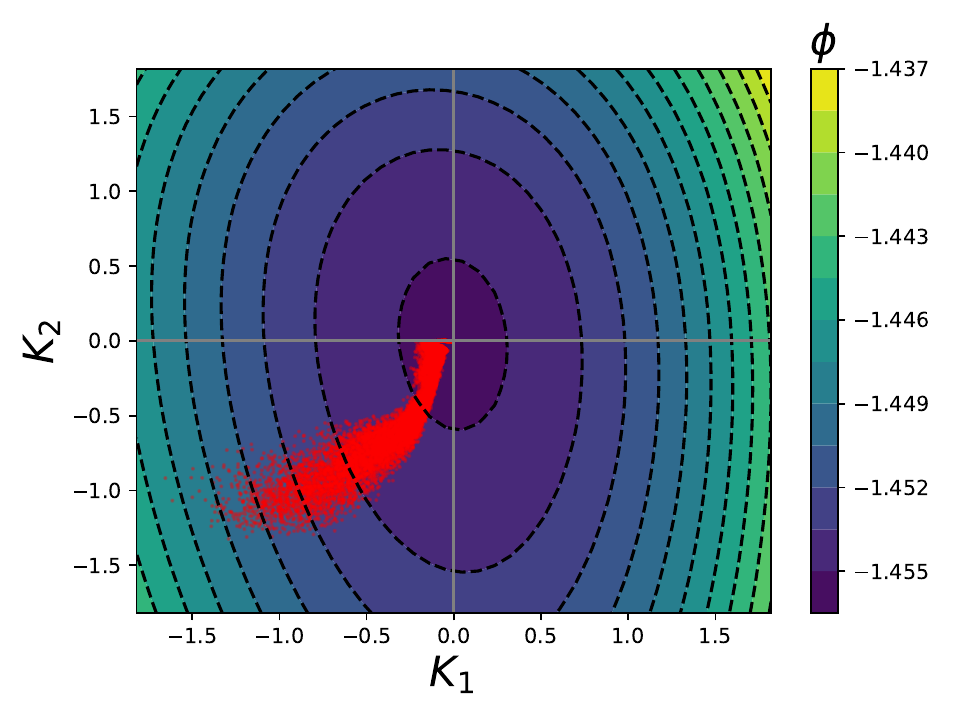}
\caption{conjugate force $K_1$ : $K_2$}
\end{subfigure}
\begin{subfigure}[c]{0.3\linewidth}
\includegraphics[width=0.9\linewidth]{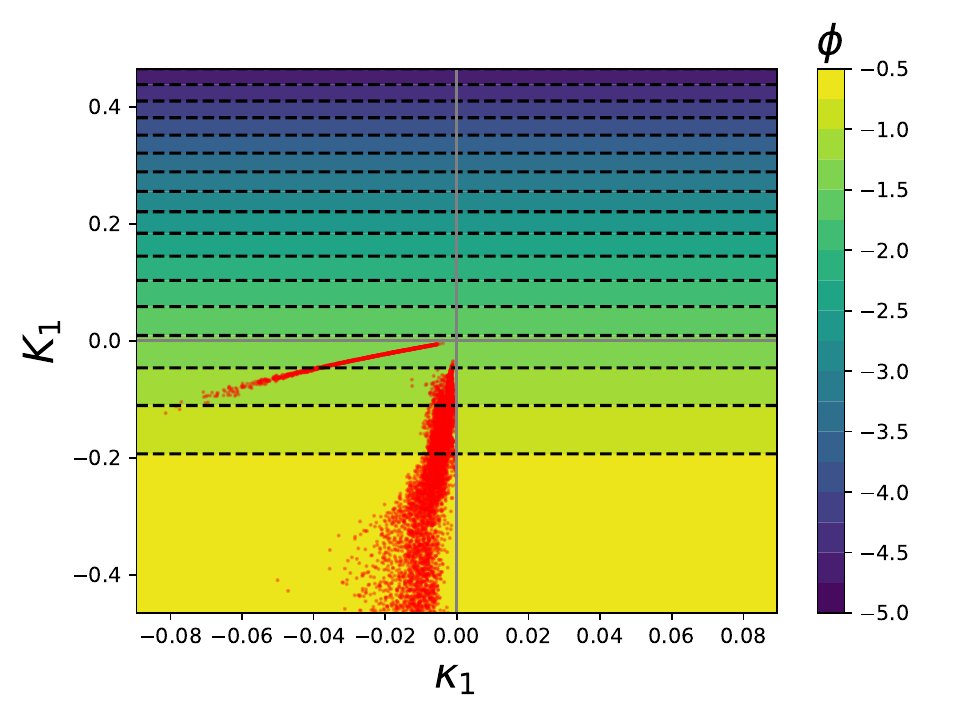}
\caption{state $\kappa_1$ : conjugate force $K_1$}
\end{subfigure}
\begin{subfigure}[c]{0.3\linewidth}
\includegraphics[width=0.9\linewidth]{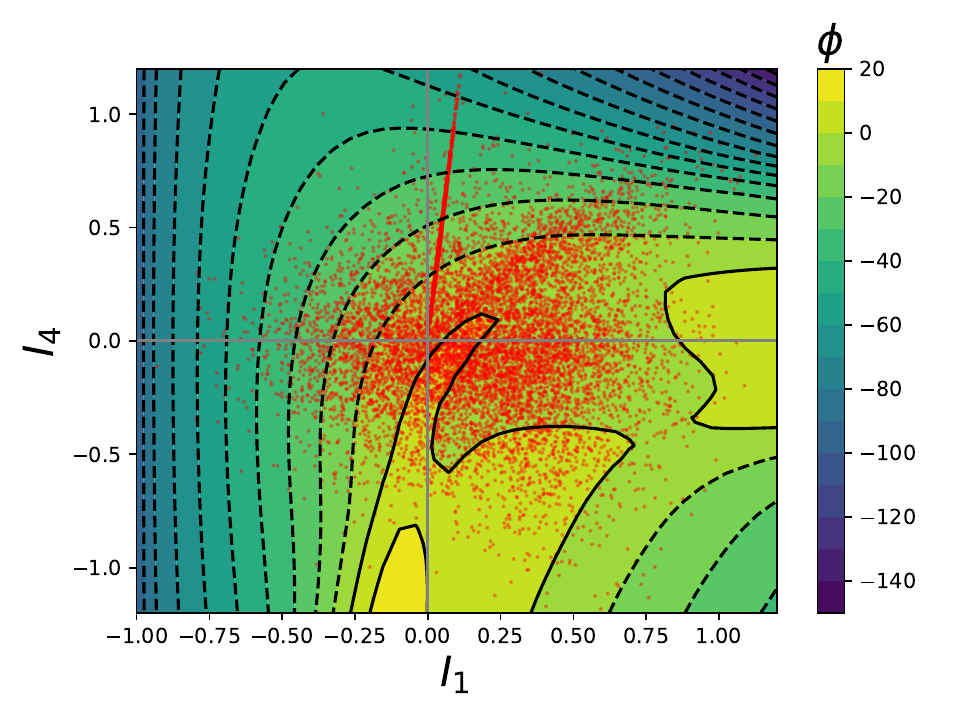}
\caption{invariant $I_1$ : $I_4$.}
\end{subfigure}

\caption{Visualization of the free energy $\psi$ and dissipation $\phi$ potentials for the nonlinear rate-dependent hardening data.
Red points are samples from the prediction trajectories.
Free energy
(a) invariant $I_1$ : invariant $I_3$,
(b) invariant $I_2$ : invariant $I_3$,
(c) internal state $\kappa_1$ : invariant $I_3$.
Dissipation potential
(d) conjugate force $K_1$ : conjugate force $K_2$,
(e) internal state $\kappa_1$ : conjugate force $K_1$,
(f) invariant $I_1$ : invariant $I_4$.
}
\label{fig:viz_vp}
\end{figure}

We also visualized projections of the free energy and dissipation potential.
\fref{fig:viz_j2} shows projections for the rate-independent PANN fit and  \fref{fig:viz_vp} shows corresponding projections for the fully viscoplastic PANN model.
For each projection, the red points indicate the regions traversed by the prediction trajectories.
Note that both the contours and the points are from the PANN, as there is no ground-truth for these quantities.
The limited locus of the accessible invariants depicted by the red trajectories has been well documented \cite{fuhg2022physics}.
It seems to indicate that different potentials can replicate the contours along these trajectories but deviate where there are no accessible invariants.

There are a number of similarities and differences between the two phenomenologies shown in \fref{fig:viz_j2} and \fref{fig:viz_vp}.
First, both PANN models exhibit broadly comparable potentials, reflecting the shared underlying elastic and plastic material properties of the data-generating models.
In fact, the $I_2$--$I_3$ free energy contours are nearly identical in \fref{fig:viz_j2}b and \fref{fig:viz_vp}b; while, the $I_1$--$I_3$ free energy contours are less similar in \fref{fig:viz_j2}a and \fref{fig:viz_vp}a, which is also plausible since the rate-independent data has a finite elastic region and the fully viscoplastic data does not.
Furthermore, panels a,b, and d are consistent with the embedded convexity of the zero-reference, centered free energy $\freeenergy$ and the centered dissipation potential $\dissipationpotential$; while the remaining panels display contours of the non-convex dependencies.
The dissipation potential for rate-independent data, \fref{fig:viz_j2}d, shows a distinctly flat bottom relative to that of the fully viscoplastic model \fref{fig:viz_vp}d; however, the regions accessed by the prediction trajectories are qualitatively similar.
In fact, in both models, there is a strong directionality to internal state-conjugate force projections in the e panels, and distinctly less directional evolution in the observable invariant projections ($I_1$:deformation, $I_4$:rate), which may be due to lower or cyclic dependencies of the dissipation potential to these inputs.

\begin{figure}
\centering
\begin{subfigure}[c]{0.95\linewidth}
\includegraphics[width=0.95\textwidth]{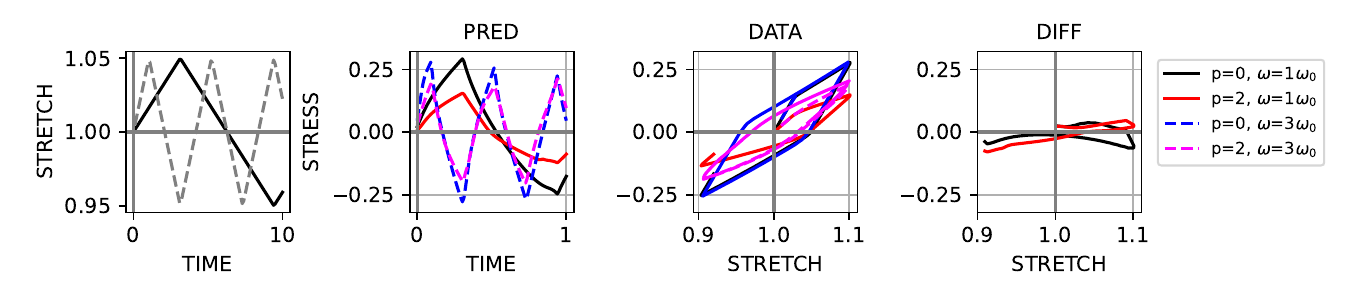}
\caption{no penalty, $\lambda_\text{KKT} = 0$}
\end{subfigure}
\begin{subfigure}[c]{0.95\linewidth}
\includegraphics[width=0.95\textwidth]{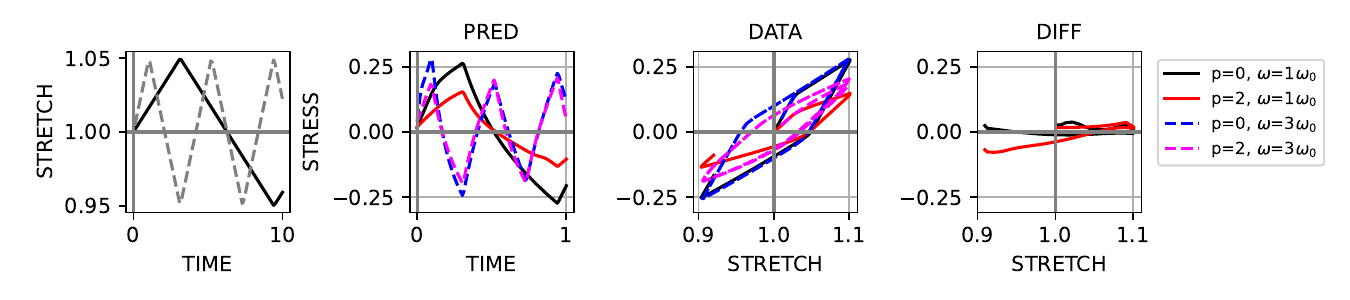}
\caption{with penalty, $\lambda_\text{KKT} =10^{-6}$}
\end{subfigure}
\caption{Regularization of the rate dependence via the KKT penalty $\lambda_\text{KKT}$: (a) with zero KKT penalty, (b) with small KKT penalty.
Fits to rate-independent ($p=0$) and rate-dependent ($p=2$) data are shown, as are fits to two different rates ($\omega_0$ and $3\omega_0$); (left to right)  strain loading at two different rates, corresponding NN response as a function of time, hysteresis loops for the data source, hysteresis difference for data with the same rate dependence ($p$) at different rates.
Note that the right-most panel shows the hysteresis error between a sawtooth trajectories that only differ by their frequencies $\{ 1, 3\} \omega_0$, i.e. there should be no difference in the hysteresis for a rate-independent model ($p=0$ data). }
\label{fig:kkt_comp}
\end{figure}

\begin{figure}
\centering
\includegraphics[width=0.45\linewidth]{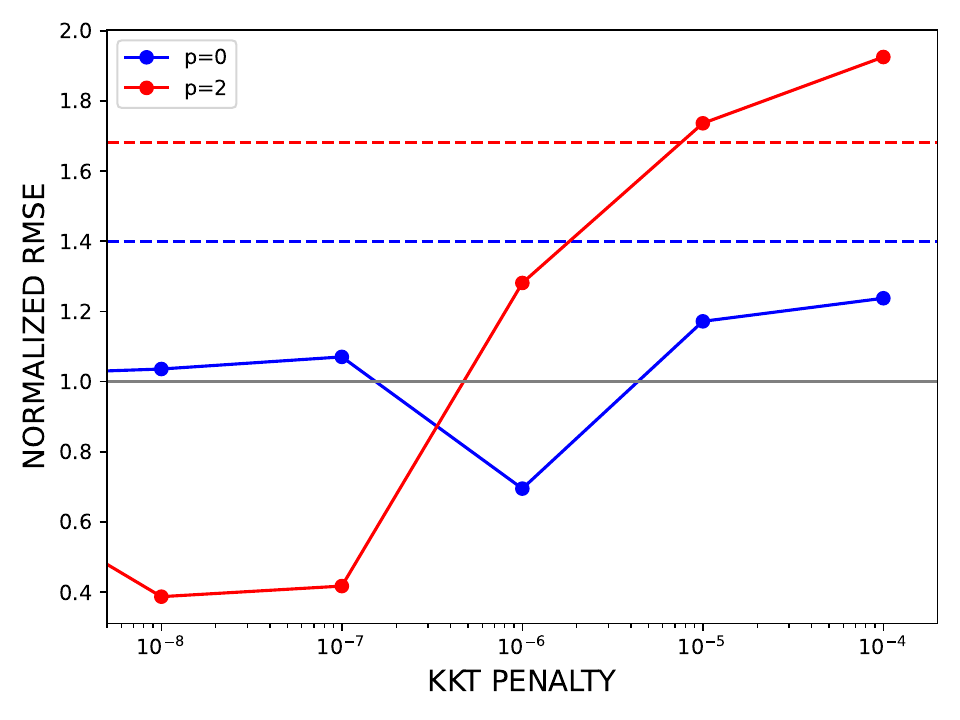}
\caption{Regularization of the rate dependence with KKT penalty $\lambda_\text{KKT}$ in loss \eqref{eq:loss}. Errors normalized by RMSE with $\lambda_\text{KKT}=0$ and horizontal dashed lines are the errors for the corresponding model without deformation rate dependence in the dissipation potential.
}
\label{fig:kkt_error}
\end{figure}

Lastly, we used a set of Cowper-Symmonds \cite{cowper1957strain} models with different exponents on rate dependence (refer to \eqref{eq:rate_dep_hardening} and \eqref{eq:CS}) to explore the effect of the KKT regularization on the loss \eqref{eq:loss}.
With the elastic region penalty fixed at $\lambda_e = 10^{-4}$ (as in the previous trials), we varied the KKT penalty $\lambda_\text{KKT}$.
Rate dependence can be measured by the difference in hysteresis between two different time parameterizations of the same cyclic deformation:
\begin{equation} \label{eq:hysteresis_diff}
\oint \pkstress(\lagrangestrain) \, \mathrm{d}\lagrangestrain(at) - \oint \pkstress(\lagrangestrain) \, \mathrm{d}\lagrangestrain(t) \, ,
\end{equation}
where $a$ is an arbitrary scalar and the path $\lagrangestrain = \lagrangestrain(t)$ is the same in both integrals.
If the model is rate independent, this difference will be zero.
\fref{fig:kkt_comp} illustrates results for fits to a rate-independent ($p=0$) and a rate-dependent ($p=2$) dataset; fits to a rate-dependent dataset are shown for comparison of the deviations.
The hysteresis difference, \eref{eq:hysteresis_diff} with $a=3$, for two different rates with the same deformation path is shown in the right panels, and
is minimized with a mild $\lambda_\text{KKT} = 10^{-6}$  (lower panels) relative to a fit with no KKT penalty (upper panels).
In fact, the errors shown in \fref{fig:kkt_error} illustrate, that even for rate-dependent data $p=0$ data, there is a positive regularizing effect (across a 200-sample testing set) that we attribute to stabilizing weights that are not well-informed by the data.
Furthermore, the results in \fref{fig:kkt_error} show that a dissipation potential without the observable rate-dependent invariants $\{ I_a\}_{a=4}^7$ produces a reasonable but inferior model to the other models that include this information in $\dissipationpotential$.

\begin{table}[h]
\centering
\begin{tabular}{|l|c|c|}
\hline
\textbf{Type} & \textbf{sine}  & \textbf{sawtooth} \\
\hline
\hline
rate-independent                         & 0.0183 & 0.0172 \\
rate-dependent linear hardening          & 0.0258 & 0.0230 \\
rate-dependent hardening                 & 0.0235 & 0.0196 \\
viscoplasticity                          & 0.0322 & 0.0275 \\
\hline
\end{tabular}
\caption{Viscoplastic exemplars: root mean squared errors (RMSEs) on held out data.
sine: in-distribution relative to the training data, sawtooth: out-of-distribution.
}
\label{tab:viscoplastic_errors}
\end{table}

\subsection{Polycrystalline stochastic volume homogenization} \label{sec:poly}

For training/testing data, we ran 400 sinusoidal loading paths over 4 temperatures $T = \{ 100, 300, 500, 700 \} K$ with a standard finite element method \cite{roters2010crystal}, of which 334 completed successfully.
Each full stress-strain response was an average over a small number of realizations (3 polycrystals) to promote isotropy.
This dataset was split into 70/20/10 training/validation/testing sets.

\begin{figure}
\centering
\includegraphics[width=0.45\linewidth]{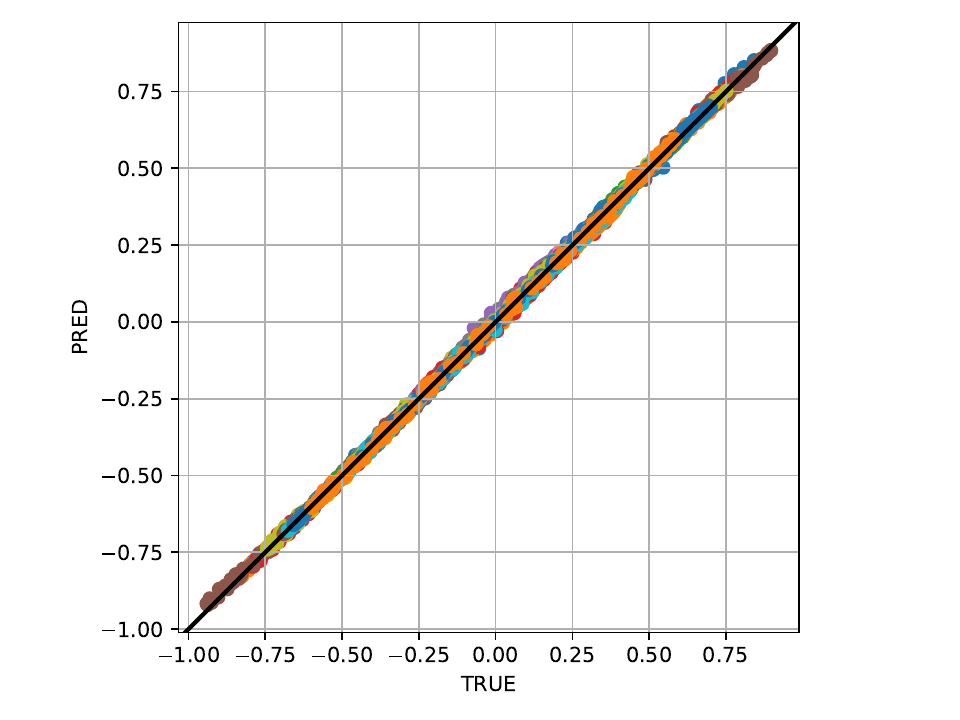}
\caption{Crystal plasticity: model parity on held-out data (colored by trajectory) with RMSE 0.0044.}
\label{fig:cp_correlation}
\end{figure}

\begin{figure}
\centering
\begin{subfigure}[c]{0.32\linewidth}
\includegraphics[width=0.99\linewidth]{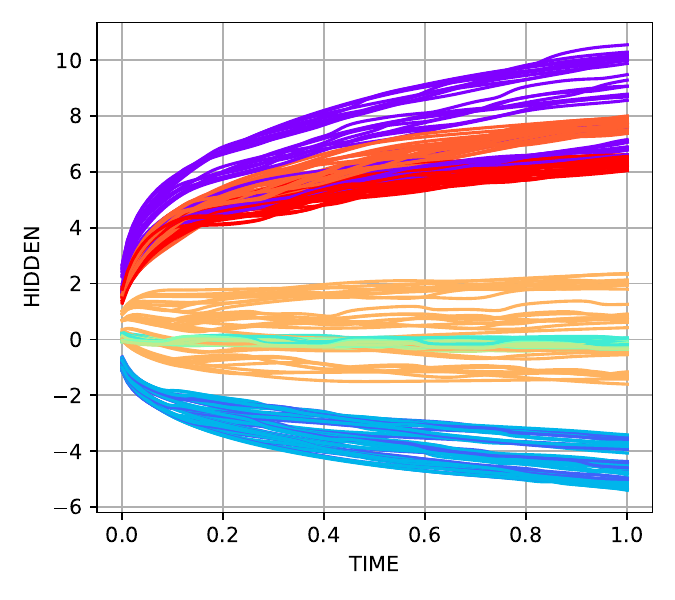}
\caption{internal state}
\end{subfigure}
\begin{subfigure}[c]{0.32\linewidth}
\includegraphics[width=0.99\linewidth]{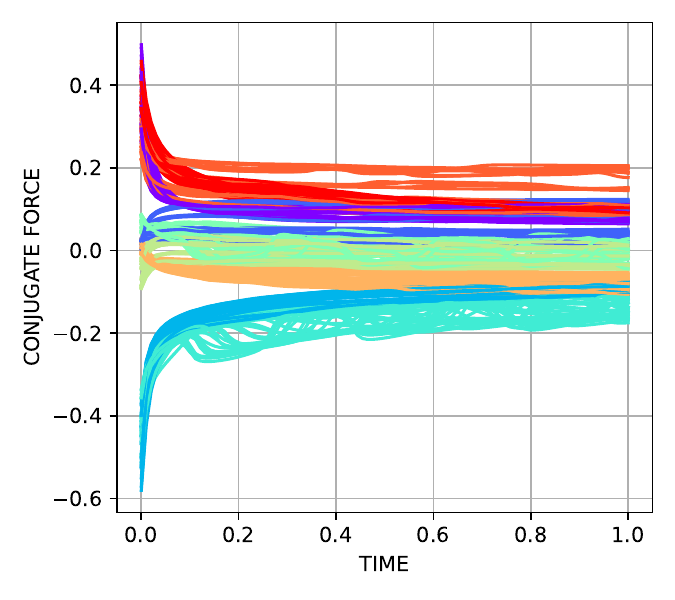}
\caption{conjugate force}
\end{subfigure}
\begin{subfigure}[c]{0.32\linewidth}
\includegraphics[width=0.99\linewidth]{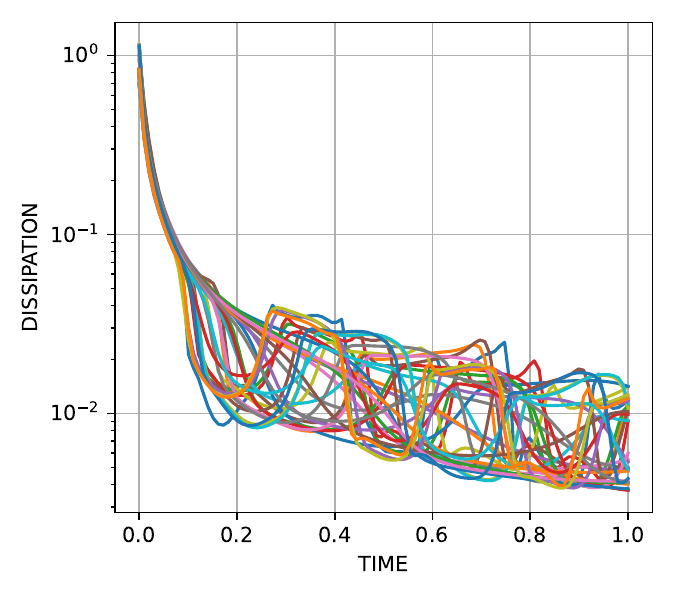}
\caption{dissipation}
\end{subfigure}
\caption{Crystal plasticity: (a) internal state $\internalvariables$, (b) conjugate force $\conjugateforce$, and (c) dissipation $\dissipation$ trajectories.
Both (a) and (b) are colored by component (multiple realizations share these colors), while (c) is colored by realization.
}
\label{fig:cp_dissipation}
\end{figure}

\fref{fig:cp_correlation} shows that, despite the complicated stress paths shown in \fref{fig:cp_hysteresis}, the predictions on held-out data, are accurate.
Furthermore, \fref{fig:cp_dissipation} illustrates the internal state and conjugate force trajectories, and the resulting dissipation evolution.
The roughly exponential increase in the components of the internal states $\internalvariables$, the decay of the conjugate force components $\conjugateforce$, and the oscillatory decay of the dissipation $\dissipation$ are consistent with the fact that sinusoidal loading will set up a non-equilibrium steady state in the SVEs.

\subsection{Tension experiment}
\begin{figure}
\centering
\includegraphics[width=0.5\linewidth]{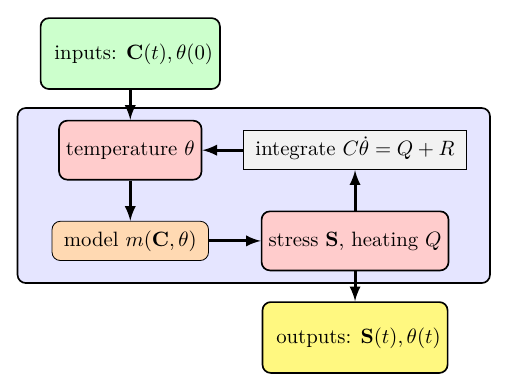}
\caption{Tension: integration scheme. Note for the data model $Q = Q_p$ in \eref{eq:tension_heat_equation} and for the NN model $Q = \dissipation$ from \eref{eq:internal_dissipation}.}
\label{fig:bvp}
\end{figure}

\begin{figure}
\centering
\includegraphics[width=0.75\linewidth]{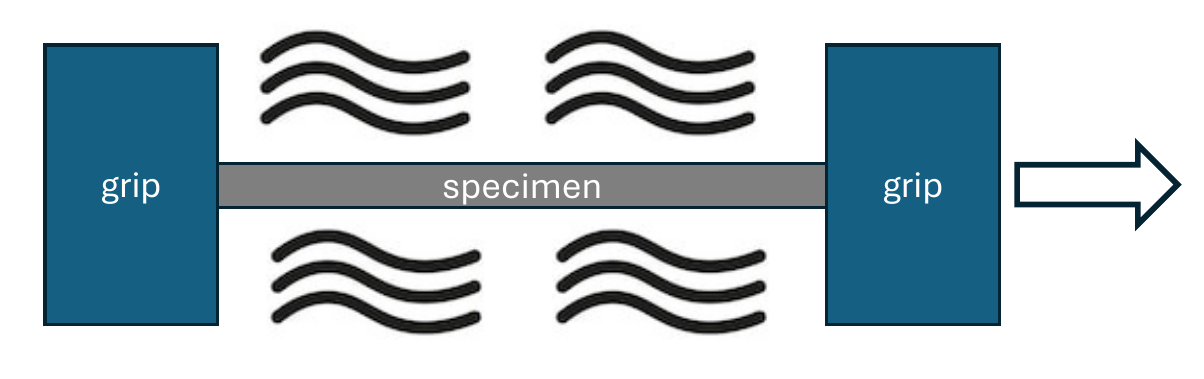}
\caption{Tension experiment: schematic. Specimen is thin and in a uniform convective atmosphere, while the grips are insulated. }
\label{fig:tension}
\end{figure}

For this demonstration, we ran two tension loading cases at various frequencies and amplitudes, and measure the resulting stress and temperature.
Since the specimen gauge is thin and the grips are insulated (refer to \fref{fig:tension}), we assume that the stress and temperature states are spatially homogeneous.
We integrate the heat equation \eqref{eq:tension_heat_equation} over the gauge section $\Omega$ to obtain
\begin{equation}
C \dot{\temperature} = Q + R\, .
\end{equation}
Here, $C = \int_\Omega \density c_\strain\, \mathrm{d}V$.
Furthermore, we set $Q=Q_p$ from \eref{eq:tension_heat_equation} for the data generating model and $Q=\dissipation$ in \eref{eq:dissipation} for the NN. In both cases, $R = \zeta \theta A$ where $\zeta$ is the cooling coefficient and $A$ is the exposed surface area of the gauge.
Recall $\temperature$ is zero at the ambient temperature.
\fref{fig:bvp} illustrates the integration scheme.

\begin{figure}
\centering
\begin{subfigure}[c]{0.65\linewidth}
\includegraphics[width=0.95\linewidth]{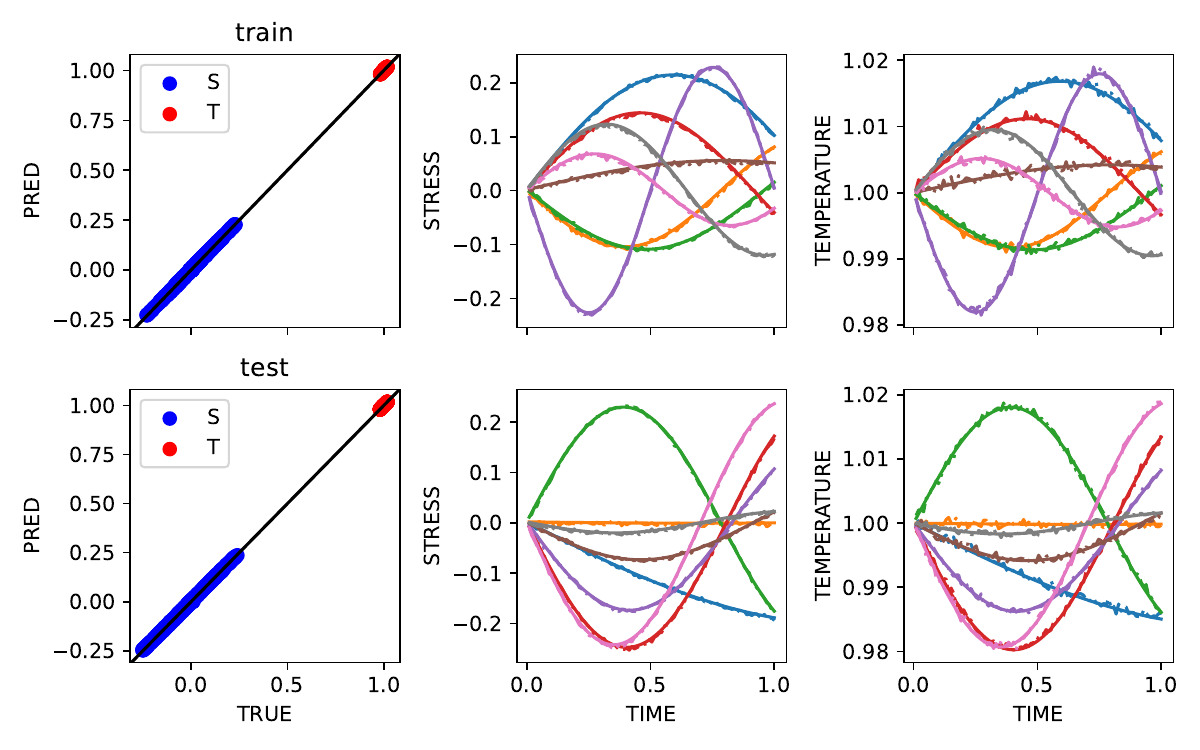}
\caption{elastic}
\end{subfigure}
\begin{subfigure}[c]{0.65\linewidth}
\includegraphics[width=0.95\linewidth]{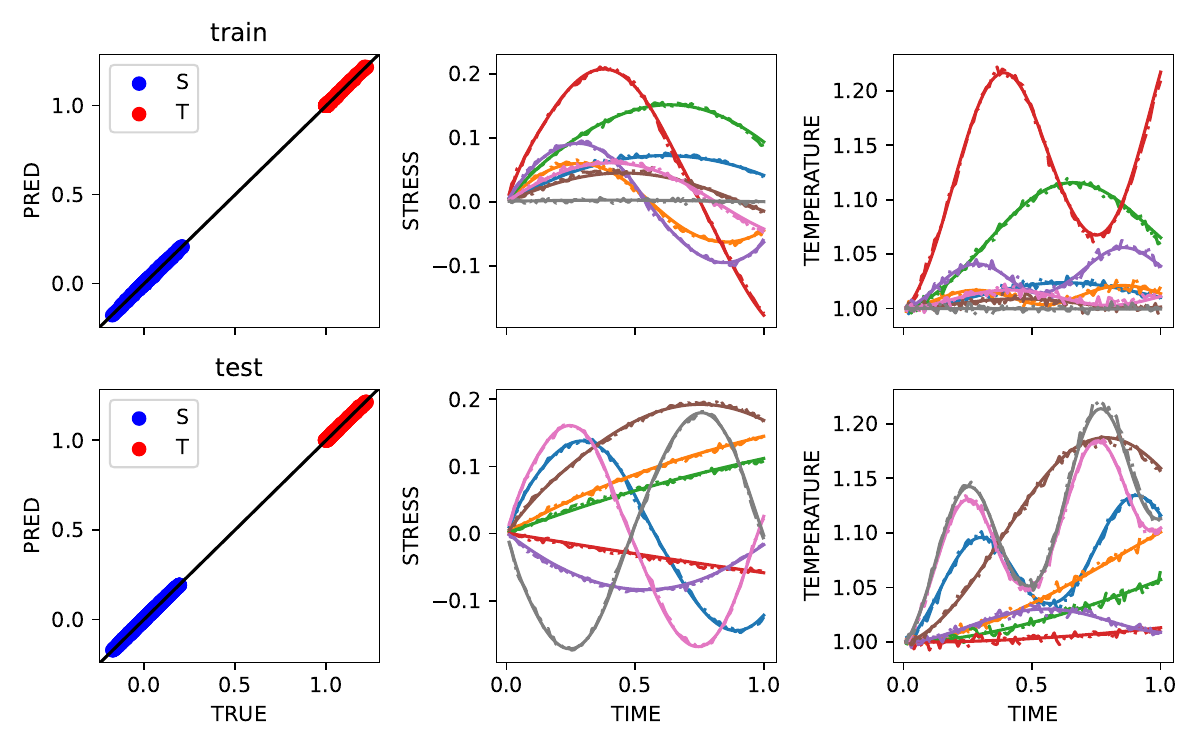}
\caption{plastic}
\end{subfigure}
\caption{Tension experiment: comparison of predictions with held out (noisy) data for: (a) thermoelastic, (b) thermoviscoplastic cases.
Note 5\% Gaussian, mean-zero noise is added to both the stress and temperature outputs.}
\label{fig:tension-evolution}
\end{figure}

We consider two cases: thermoelastic and thermoviscoplastic, where we observe stress and temperature with measurement noise (additive Gaussian white noise with standard deviation $\varsigma$) (parameters are given in \tref{tab:tension_data}).
In both cases, the same PANN architecture displays excellent fidelity with the held-out data shown in  \fref{fig:tension-evolution}.
Note for the thermoelastic case, the temperature evolves around the ambient temperature since the thermoelastic material has no bias in heating in compression (or cooling in expansion).
On the other hand, with the thermoplastic material, the heating due to plastic work dominates the heat sink due to hardening.
The quantitative RMSE for held-out data for both cases is summarized in \tref{tab:tension}.

\begin{table}[h]
\centering
\begin{tabular}{|lc|cc|}
\hline
\textbf{Parameter} & & \textbf{thermoelastic} & \textbf{viscoplastic} \\
\hline
Bulk modulus          & $K$           & 1.0 & 1.0 \\
Shear modulus         & $G$           & 0.4 & 0.4 \\
Thermal expansion coefficient & $\thermalexpansioncoefficient$ & 0.05 & 0.001 \\
\hline
Heat capacity         & $\heatcapacity$                & 0.1 & 0.1 \\
Reference temperature & $\abstemperature_\text{ref}$            & 1.0 & 1.0 \\
Cooling coefficient   & $\zeta$                                & 0.001 & 0.001 \\
\hline
Initial yield         & $Y_0$                                  & - & 0.02 \\
Saturation yield      & $Y_\infty$                             & - & 0.025 \\
Reference plastic strain & $Y_\infty$                          & - & 0.1 \\
Viscous relaxation time  & $\tau$                              & - & 0.1 \\
\hline
Measurement noise     &  $\varsigma$                           & 0.05 & 0.05 \\
\hline
\end{tabular}
\caption{Tension data model parameters}
\label{tab:tension_data}
\end{table}

\begin{table}[h]
\centering
\begin{tabular}{|cc|cc|cc|}
\hline
&       & \multicolumn{2}{c|}{\textbf{Training}} &   \multicolumn{2}{c|}{\textbf{Testing}} \\
\textbf{Type} &   \textbf{N} &  \textbf{stress} & \textbf{temperature} & \textbf{stress} & \textbf{temperature}  \\
\hline
\hline
Elastic &   100 & 0.00323 & 0.00041 & 0.00341 & 0.00041 \\
&   512 & 0.00249 & 0.00040 & 0.00248 & 0.00040 \\
\hline
Plastic &   100 & 0.00470 & 0.00470 & 0.00555 & 0.00561 \\
&   512 & 0.00248 & 0.00313 & 0.00247 & 0.00321 \\
\hline
\end{tabular}
\caption{Tension experiment: RMSE on training and held-out testing data with noise with increasing data.}
\label{tab:tension}
\end{table}

\section{Conclusion} \label{sec:summary}

In order to construct a universal model for rate-independent and rate-dependent thermoplasticity, we designed a learning framework with functional relationships between the free energy and dissipation potential akin to the theory developed by Chaboche\cite{chaboche1987continuum,chaboche2001unified,lemaitre1994mechanics}.
No duality nor Legendre-Fenchel transforms are needed, and, hence, we assert that the framework is less restrictive than most GSM models.
In addition, the presence of deformation and rate measures in the dissipation potential is not proscribed and was shown to help the predictive capacity of the chosen framework.
It is also distinct from our previous work \cite{jones2025attention}, which utilized a gated flow to learn elastic versus inelastic behavior, on two key points: (a) the employment of a convex dissipation potential to ensure second law dissipation requirements, and (b) KKT conditions in the loss to sharpen the flow transitions toward a yield-like gate.
The performance and utility of the framework were illustrated with demonstrations ranging across: (a) representation of inviscid elastic-plastic to fully viscoplastic flow, (b) homogenization of the complex behavior of SVEs, and (c) a mock calibration to tension experiment with significant heat generation.

In future work, we intend to pursue multiple avenues.
First, we will adapt the model to elastic-plastic sequential training that may have advantages for complex, noisy, low-count experimental data and include hybrid models for the elastic, thermal, and inelastic aspects \cite{fuhg2023modular,fuhg2024polyconvex}.
Along these lines, we expect sparsification of models generated by the proposed framework in terms of parameters \cite{fuhg2023extreme,padmanabha2024improving,padmanabha2025concurrent}, inputs \cite{fuhg2022learning,patel2025general}, and components \cite{fuhg2023modular} will be beneficial.
This will likely include winnowing observable invariants and possibly extending joint internal-observable invariant measures.

\section*{Acknowledgements}
This work was funded by the Laboratory Directed Research and Development (LDRD) program at Sandia National Laboratories; this funding is gratefully acknowledged.
Sandia National Laboratories is a multi-mission laboratory managed and operated by National Technology \& Engineering Solutions of Sandia, LLC (NTESS), a wholly owned subsidiary of Honeywell International Inc., for the U.S. Department of Energy’s National Nuclear Security Administration (DOE/NNSA) under contract DE-NA0003525. This written work is authored by an employee of NTESS. The employee, not NTESS, owns the right, title and interest in and to the written work and is responsible for its contents. Any subjective views or opinions that might be expressed in the written work do not necessarily represent the views of the U.S. Government. The publisher acknowledges that the U.S. Government retains a non-exclusive, paid-up, irrevocable, world-wide license to publish or reproduce the published form of this written work or allow others to do so, for U.S. Government purposes. The DOE will provide public access to results of federally sponsored research in accordance with the DOE Public Access Plan. \\
This material is based upon work partially supported by the U.S. National Science Foundation under award No. 2452029.
The opinions, findings, and conclusions, or recommendations expressed are those of the authors and do not necessarily reflect the views of the NSF.
The authors acknowledge the Texas Advanced Computing Center (TACC) at The University of Texas at Austin for providing computational resources that have contributed to the research results reported within this paper.



\appendix
\numberwithin{equation}{section}

\section{Viscoplastic exemplars} \label{app:vp_exemplars}

Each model has three components: (a) a stress rule, (b) a yield condition, and (c) a flow rule.

The stress component is the same for all exemplars, where we use a  linear finite elastic St. Venant model \cite{horgan1981effect}:
\begin{equation}
\pkstress
= K \, \tr \lagrangestrain_e \Ib + 2 G \, \dev \lagrangestrain_e \ ,
\end{equation}
for stress ($K$: bulk modulus, $G$: shear modulus) and additive decomposition of the strain measures  \cite{lehmann1984general,miehe1995theory,naghdi1990critical}:
\begin{equation} \label{eq:additive_split}
\lagrangestrain_e = \lagrangestrain - \lagrangestrain_p \ ,
\end{equation}
which eases the translation of the algorithms in \cref{simo2006computational} to the finite deformation context.
The yield condition allows for isotropic hardening:
\begin{equation}
\yield(\pkstress,\internalvariables) = \| \dev \pkstress \|_2 - h(\eqps, \doteqps, \temperature ) \le 0 \ ,
\end{equation}
with the cumulative scalar invariant: equivalent plastic strain $\eqps$.
Associative flow leads to:
\begin{equation}
\dot{\lagrangestrain}_p = \partialb_\pkstress \yield = \consistencyrate \Nb = \consistencyrate \frac{\dev \pkstress}{\| \dev \pkstress \|}
\end{equation}
where $\consistencyrate$ is a constitutive function $\consistencyrate = \consistencyrate(\eqps, \doteqps, \temperature, \ldots)$ for the rate-dependent/viscoplastic classes.
Here $\eqps$ is the scalar accumulated plastic strain measure and $\temperature$ is the empirical temperature.
This function differs qualitatively between the classes.

\paragraph{Rate-independent}
For this exemplar, we utilize the von Mises/J2 model with a finite elastic region and
linear hardening $h = Y + H\epsilon_p$:
\begin{eqnarray}
\yield &=& \| \dev \pkstress \| - (Y + H\epsilon_p)  \\
&=& 2 G \| \dev \lagrangestrain_e \| - (Y + H\epsilon_p) \ ,  \nonumber
\end{eqnarray}
where $Y$ is initial yield, and $H$ is hardening modulus.
To solve for equilibrium in this case a trial stress $\dev \pkstress = 2 G \, \dev \lagrangestrain^\text{trial}_e + \consistencyparameter \Nb$ and return map:
\begin{equation}
\consistencyrate \dt = \Delta\consistencyparameter = \frac{\langle \yield \rangle}{2 G + H } \equiv \Delta \eqps \ ,
\end{equation}
are employed.

\paragraph{Rate-dependent linear hardening}
This exemplar is similar to the rate-independent one just described, except the material is allowed to exceed $\yield = 0$.
When $\yield > 0$ it relaxes to $\yield=0$ via:
\begin{equation}
\dot{\gamma} = \frac{1}{\eta} \langle \yield \rangle \ ,
\end{equation}
where $\tau = \tfrac{\eta}{2 G}$ is the associated time scale.
In this case, the update is:
\begin{equation}
\Delta \gamma = \frac{\langle\yield\rangle}{2 \mu + H + \eta / \dt } \ .
\end{equation}

\paragraph{Rate-dependent hardening}
For this case, we employ the well-known Johnson-Cook \cite{johnson1983constitutive} separable hardening rule:
\begin{equation} \label{eq:rate_dep_hardening}
h(\eqps,\doteqps) = h_{\eqps}(\eqps) \, h_{\doteqps} (\doteqps) =
\left(a + b \eqps^n\right)
\left(1 + c \, \log\frac{\doteqps}{\dot{\epsilon}_0} \right) \ ,
\end{equation}
where $a$, $b$, $c$, $\dot{\epsilon}_0$ are material parameters.
Alternately, the Cowper-Symmonds \cite{cowper1957strain} power law:
\begin{equation} \label{eq:CS}
h_{\doteqps} =
1 + c \left(\frac{\dot{\epsilon}_p}{\dot{\epsilon}_0} \right)^p \ ,
\end{equation}
is employed in additional investigations because the degree of rate dependence can be controlled by the parameter $n$.
Since these are nonlinear, a root finding method is needed to determine the growth of the yield surface via $\dot \gamma$.

\paragraph{Pure viscoplasticity}
For this exemplar, we use an overstress formulation  with a negligible elastic region:
\begin{equation}
\yield = \| \dev \pkstress \| \ ,
\end{equation}
and
\begin{equation}
\dot{\gamma} = \frac{1}{\eta} \left( \frac{ \langle \yield \rangle} {\sigma_y} \right)^p \ ,
\end{equation}
where $\eta$ is a viscosity, $p$ the exponent in power law, and $\sigma_y$ is a strength parameters.

\begin{table}[h]
\centering
\begin{tabular}{|l|c|c|}
\hline
{\bf Rate-independent linear hardening} & & \\
\hline
Young's modulus & $E$ & 2.0 \\
Poisson's ratio & $\nu$ & 0.3 \\
(initial) yield & $Y$   & 0.02 \\
hardening       & $H$   & 0.05 \\
\hline
{\bf Rate-dependent linear hardening} & & \\
relaxation time     & $\eta$ & 0.4 \\
\hline
{\bf Rate-dependent hardening} & & \\
\hline
hardening constant    & $a$ &  0.02 \\
hardening coefficient & $b$ &  0.05 \\
hardening exponent    & $n$ & 0.5 \\
hardening rate coefficient & $c$ &  5.5 \\
hardening rate reference   & $\dot{\epsilon}_0$ &  0.1 \\
\hline
{\bf Viscoplastic} & & \\
\hline
hardening rate constant & $\tau$ & 1.0 \\
viscosity exponent      & $p$    & 1.5 \\
yield         & $\sigma_Y$   & 0.001 \\
\hline
\end{tabular}
\caption{Viscoplastic exemplars: parameters are non-dimensionalized for comparison.
For brevity, only the parameters that differ from the preceding models are listed.}
\label{tab:stvenant}
\end{table}

\FloatBarrier

\section{Polycrystalline model} \label{app:sve}

Within a realization of a polycrystalline aggregate, the response of each crystal is given by an elastic-viscoplastic constitutive relation based on well-known meso-scale models of single crystal deformation  \cite{taylor1934mechanism,kroner1961plastic,bishop1951xlvi,bishop1951cxxviii,mandel1965generalisation,dawson2000computational,roters2010overview} using parameters representative of face-centered cubic (FCC) Fe given in \tref{tab:cp_Fe}.
For the crystal elasticity, we employed a linear relation between the second Piola-Kirchhoff stress $\Sb$ in the intermediate configuration and the elastic Lagrangian strain $\Eb^e$,
\begin{equation}
\Sb = \Cbb : \Eb^e \ .
\end{equation}
For a FCC crystal, the independent components of the elastic modulus tensor $\Cbb$ are $C_{1111}, C_{1122}, C_{1212}$.
In each crystal, plastic flow can occur on any of the 12 FCC slip planes, where we employed an Arrhenius-like slip rate relation
\begin{equation}
\dot{\gamma} = \dot{\gamma}_0 \left(\frac{\tau}{g_\alpha}\right)^{m_0 \frac{k_B \abstemperature_0}{k_B \abstemperature}}
\end{equation}
driven by the shear stress $\tau_\alpha$ resolved on slip system $\alpha$.
Here, $\dot{\gamma}_0$ is the reference slip rate, and $g_{\alpha}$ is the slip resistance which evolves according to \cite{Kocks1976, mecking1976hardening}:
\begin{equation}
\dot{g}_\alpha = (H-R g_\alpha) \sum_\alpha |\dot{\gamma}_\alpha| \ ,
\end{equation}
where $H$ is the hardening modulus and $R$ is the temperature-dependent recovery coefficient \cite{thiercelin2023thermally}:
\begin{equation}
R(\abstemperature) =  R_0 \exp\left(-\frac{\abstemperature_0}{\abstemperature} \right) \ .
\end{equation}
A homologous temperature $\abstemperature^* = \abstemperature/\abstemperature_0$ helps to simplify these relationships.

\begin{table}[]
\centering
\begin{tabular}{|lc|c|}
\hline
Elastic modulus
& C11 & 204.6 [GPa] \\
Elastic modulus
& C12 & 137.7 [GPa] \\
Elastic modulus
& C44 & 126.2 [GPa] \\
\hline
Reference slip rate &$\dot{\gamma}_0$& 1.0 [1/s]  \\
Rate exponent &$m_0$& 20.0        \\
Hardening modulus & $H$ & 600 [MPa] \\
Recovery modulus & $R_0$ & 1.0 \\
Initial slip resistance   & $g_\alpha(0)$ & 122 [MPa] \\
\hline
\end{tabular}
\caption{Crystal plasticity parameters from Refs. \cite{jones2018machine,frankel2019predicting}}
\label{tab:cp_Fe}
\end{table}
\end{document}